\documentclass[english]{article}
\usepackage[T1]{fontenc}
\usepackage[latin9]{inputenc}
\usepackage{color}
\usepackage{verbatim}
\usepackage{float}
\usepackage{amsmath}
\usepackage{amssymb}
\usepackage{stackrel}
\usepackage{stmaryrd}
\usepackage{graphicx}
\PassOptionsToPackage{normalem}{ulem}
\usepackage{ulem}
\pdfoutput=1
\makeatletter

\providecommand{\tabularnewline}{\\}

\usepackage{amsfonts,amssymb}
\usepackage{latexsym}
\usepackage{epsfig}
\usepackage{cite}
\usepackage{graphicx}
\usepackage[colorlinks,linkcolor=blue]{hyperref}
\newcommand{\be}{\begin{equation}}
\newcommand{\ee}{\end{equation}}

\makeatother

\usepackage{babel}
\begin{document}
{}~ \hfill\vbox{\hbox{CTP-SCU/2021037}}\break
\vskip 3.0cm
\centerline{\Large \bf Connections between reflected entropies and hyperbolic string vertices}

\vspace*{10.0ex}
\centerline{\large Peng Wang$^a$, Houwen Wu$^{a,b}$ and Haitang Yang$^a$}
\vspace*{7.0ex}
\vspace*{4.0ex}
\centerline{\large \it $^a$College of Physics}
\centerline{\large \it Sichuan University}
\centerline{\large \it Chengdu, 610065, China} \vspace*{1.0ex}
\vspace*{4.0ex}
\centerline{\large \it $^b$DAMTP, Centre for Mathematical Sciences}
\centerline{\large \it University of Cambridge}
\centerline{\large \it Cambridge, CB3 0WA, UK} \vspace*{1.0ex}
\vspace*{4.0ex}

\centerline{pengw@scu.edu.cn, hw598@damtp.cam.ac.uk, hyanga@scu.edu.cn}
\vspace*{10.0ex}
\centerline{\bf Abstract} \bigskip \smallskip
In this paper, we establish connections between the reflected entropies of  multipartite mixed states
in CFT$_{2}$ and hyperbolic string vertices of closed string field theory (CSFT). We show that
the reflected surfaces,
which are bulk duals of the  reflected entropies,
share the same Riemann surfaces with the hyperbolic string vertices.
This observation enables us to build quantitative relations between the
reflected entropies and hyperbolic  string vertices. We illustrate the connections with several examples.
Consequently,  we  propose that
spacetime structure could be  directly generated from the hyperbolic string vertices.
The  advantage
of the hyperbolic string vertices approach is that we have a dynamical equation, the
Batalin-Vilkoviski master equation,
to control the generating process.


\vfill
\eject
\baselineskip=16pt
\vspace*{10.0ex}
\tableofcontents

\section{Introduction}

It is well known that   closed string field theory (CSFT) has two equivalent
descriptions \cite{Hata:1993gf}. The first one uses a conformal field
theory to represent a string background. The second one adopts background
independent string vertices \cite{Sen:1993kb,Sen:1994kx}, and it therefore attracted
much more attention. In early works \cite{Zwiebach:1992ie}, Zwiebach
demonstrated that, if all consistent string vertices and
their corresponding suitable conformal field theory are known, bosonic closed
string field theory can be constructed. These consistent string
vertices are required to satisfy a geometric version of the Batalin-Vilkoviski
(BV) master equation, i.e. geometric master equation. Therefore, one
main task in CSFT is to seek all the consistent string vertices which
exactly solve the geometric master equation.

The development of the
string vertices has mainly experienced  three stages. The first approach is to
consider Riemann surfaces endowed with a minimal area metric, which enables
a decomposition of a moduli space \cite{Zwiebach:1992ie,Zwiebach:1990nh}.
It is then simple to specify the string vertices respecting the BV equation.
However, this approach only works perfectly for genus zero surfaces.
For higher genus ($g\ge 1$) surfaces, the existence of the  minimal area metric
has no proof.

As an alternative proposal, Moosavian and Pius used the  hyperbolic surfaces
to construct the string vertices \cite{Moosavian:2017qsp,Moosavian:2017sev}. The
proof of existence is no question, but the BV equation is not exactly solved.
To satisfy the BV equation, higher order corrections are needed  and their existence is unknown.

In a recent work \cite{Costello:2019fuh}, Costello and Zwiebach   made a simple
but  brilliant improvement on the Moosavian-Pius's approach,
that is,  replace the horocycle, which is
the boundary of the coordinate disks around the punctures in the Moosavian-Pius surface,
by geodesics  of length $0<L\le 2  \mathrm{arc}\sinh(1)$.
The consistent string vertices, called as hyperbolic string vertices,
are those  whose  systole\footnote{The systole $sys[\Sigma]$ of a surface $\Sigma$
is defined as the length of the shortest non-contractible closed geodesic
which is not a boundary component.}
is not less than $L$.
These hyperbolic string vertices solve the BV equation exactly.
Cho soon   generalized
this method to construct the open-closed string vertices \cite{Cho:2019anu}.

The minimal area approach and Moosavian-Pius construction turn out to be
limits of the hyperbolic string vertices.
As verified in \cite{Firat:2021ukc},
the simplest hyperbolic
$3$-string vertex  (Y-piece) with boundary length $L$,
when $L\rightarrow\infty$,   reduces to the minimal area $3$-string vertex.
On the other hand, as $L\rightarrow 0$, it becomes the naive three-string vertex
\cite{Moosavian:2017qsp,Moosavian:2017sev} or Kleinian vertex \cite{Sonoda:1989sj}.


Another background ingredient needed in this article is the reflected entropy,
which is a generalization of the entanglement entropy to the mixed state systems.
The  entanglement entropy (EE) measures
the correlation between subsystems of a pure system. It is one of the most distinct features
of quantum systems. Dividing a pure system into two subsystems: $A$ and $B$, the total Hilbert
space is accordingly decomposed as $\mathcal{H}=\mathcal{H}_{A}\otimes\mathcal{H}_{B}$.
Tracing out degrees of freedom of the region $B$, one obtains a reduced
density matrix of the region $A$: $\rho_{A}=\mathrm{Tr}_{\mathcal{H}_{B}}\rho$.
The entanglement entropy of the region $A$ is evaluated by the von
Neumann entropy $S_{A}=-\mathrm{Tr}_{\mathcal{H}_{A}}\left(\rho_{A}\log\rho_{A}\right)$.
It is clear that $S_{A}=S_{B}$.

Motivated by AdS/CFT correspondence
and Bekenstein-Hawking entropy of black holes, Ryu and Takayanagi
(RT) proposed that the minimal surface area ending on the $d$ dimensional
boundary of AdS$_{d+1}$ corresponds to the entanglement entropy of CFT$_{d}$
living on the boundary of AdS$_{d+1}$ \cite{Ryu:2006bv}. For $d=2$,
the minimal surfaces are geodesics and the RT formula has been verified
extensively.

However, the entanglement entropy is ill-defined for mixed states.
One obvious feature is that  the entanglement entropy of mixed states is always
nonzero whether they are entangled or not. One method to solve
this problem is to purify the mixed state and then
introduce the entanglement of purification $E_{P}$,
whose bulk dual is  conjectured as the area of entanglement wedge cross-section
$E_{W}$, namely $E_{P}=E_{W}$ \cite{Takayanagi:2017knl}. Usually,
calculating $E_{P}$ is very difficult  since it involves minimization
over all possible purifications.
For this reason, in  \cite{Dutta:2019gen}, a much simpler alternative has been proposed,
namely the canonical purification, which defines the reflected entropy $S_R$.
The bulk interpretation
of the canonical purification is developed in \cite{Engelhardt:2018kcs}.
The bulk dual of the reflected entropy is called reflected surface, also denoted as $S_R$ if
not confused. For a bipartite system, it turns out  $S_{R}\left(A:B\right)=2E_{W}\left(A:B\right)$.
Discussions on the multipartite reflected entropy and its bulk interpretation
can be found in refs. \cite{Bao:2019zqc,Chu:2019etd}.


Careful studies find that both the hyperbolic string vertices and
the reflected surfaces are hyperbolic surfaces bounded by geodesics.
They furthermore share very similar construction procedures.
In addition, theorems in hyperbolic geometry impose strict constraints
on the structure of surfaces.
We are therefore
motivated to establish connections between the hyperbolic
string vertices and the bulk geometries of multipartite canonical
purification.

There are two obstacles to build the connections.
The boundary lengths of the string vertices are constrained to be the same while
those of reflected surfaces are allowed to be different. We will show that this problem
can be solved by constructing procedures. A more serious problem is that
the  boundary lengths of the (quantum) string vertices have an upper bound
$L\le L_* =2\mathrm{arc}\sinh1$, but
those of reflected surfaces have a lower bound $S_R > 2L_*$.
It turns out this
sharp contradiction is solved magically by intrinsic properties of hyperbolic geometry.

Since it is widely believed that the spacetime structure could be generated by the entanglement
entropy through the dual surfaces, once the connections between the hyperbolic
string vertices and reflected entropies/surfaces are established, we are led to
ask if the spacetime  could directly emerge from the hyperbolic string vertices.
A great advantage of the  hyperbolic string vertex approach is that the generating
process is completely  controlled by the BV master equation.

This paper is organized as follows. In section 2, we provide background
reviews for hyperbolic string vertices of CSFT and canonical purification
of CFT. In section 3, we establish the connections between the closed
string vertices and the bulk geometries of canonical purification.
Several examples are supplied to illustrate the connections.
In Section 4, we provide a preliminary evidence to reinforce the connections.
In Section 5, we argue that spacetime can be built by the string vertices.
The last section includes the conclusion and discussions.

\section{Preliminaries}
In this section, we briefly review the two ingredients on which our results are based. We first show
the construction of the simplest string vertex   $\mathcal{V}_{0,3}(L)$ in detail
and give some theorems needed in the rest of this work.
Subsection $2.2$
is devoted to a simple introduction to the reflected entropy and the   dual reflected surfaces.

\subsection{Hyperbolic string vertices in CSFT}

In closed string field theory, off-shell amplitudes are defined by
a set of string vertices $\mathcal{V}_{g,n}$, which are
subsets of \emph{moduli spaces} $\mathcal{\hat P}_{g,n}$ of compact
Riemann surfaces of genus $g$ and $n$ marked points, with local
coordinates defined around those marked points up to  phases. These vertices
have  negative Euler numbers $2g+n-2>0$ and thus admit hyperbolic
metrics of constant negative Gaussian curvature. To get a consistent
quantum theory, these string vertices must satisfy the Batalin-Vilkoviski
(BV) master equation,

\begin{equation}
\partial\mathcal{V}+\hbar\triangle\mathcal{V}+\frac{1}{2}\left\{ \mathcal{V},\mathcal{V}\right\} =0,
\label{eq:BV equation}
\end{equation}

\noindent where

\begin{equation}
\mathcal{V}=\underset{g,n}{\sum}\hbar^{g}\mathcal{V}_{g,n},\qquad with\qquad\begin{cases}
n\geq3, & for\;g=0,\\
n\geq1, & for\;g=1,\\
n\geq0, & for\;g\geq2.
\end{cases}
\end{equation}
In eq. (\ref{eq:BV equation}), geometrically, $\partial$ indicates
the boundary of a moduli space. $\triangle$ denotes removing the disks
of two marked points on one Riemann surface, and then twist-sewing
the boundaries of these two disks. $\left\{ \;,\;\right\} $ stands
for removing two disks on two input Riemann surfaces respectively,
and then twist-sewing them together \cite{Sen:1993kb,Sen:1994kx}.
In a recent remarkable work \cite{Costello:2019fuh},
Costello and  Zwiebach   proved that the
closed string vertices can be simply and elegantly constructed by using the hyperbolic
geometry. For the purpose to show the connections  with the reflected surfaces plainly,
we demonstrate the construction of the fundamental building block $\mathcal{V}_{0,3}(L)$
from  the very beginning, in four steps:

\subsubsection*{\uline{Step one:}}

\begin{figure}[H]
\begin{centering}
\includegraphics[scale=0.7]{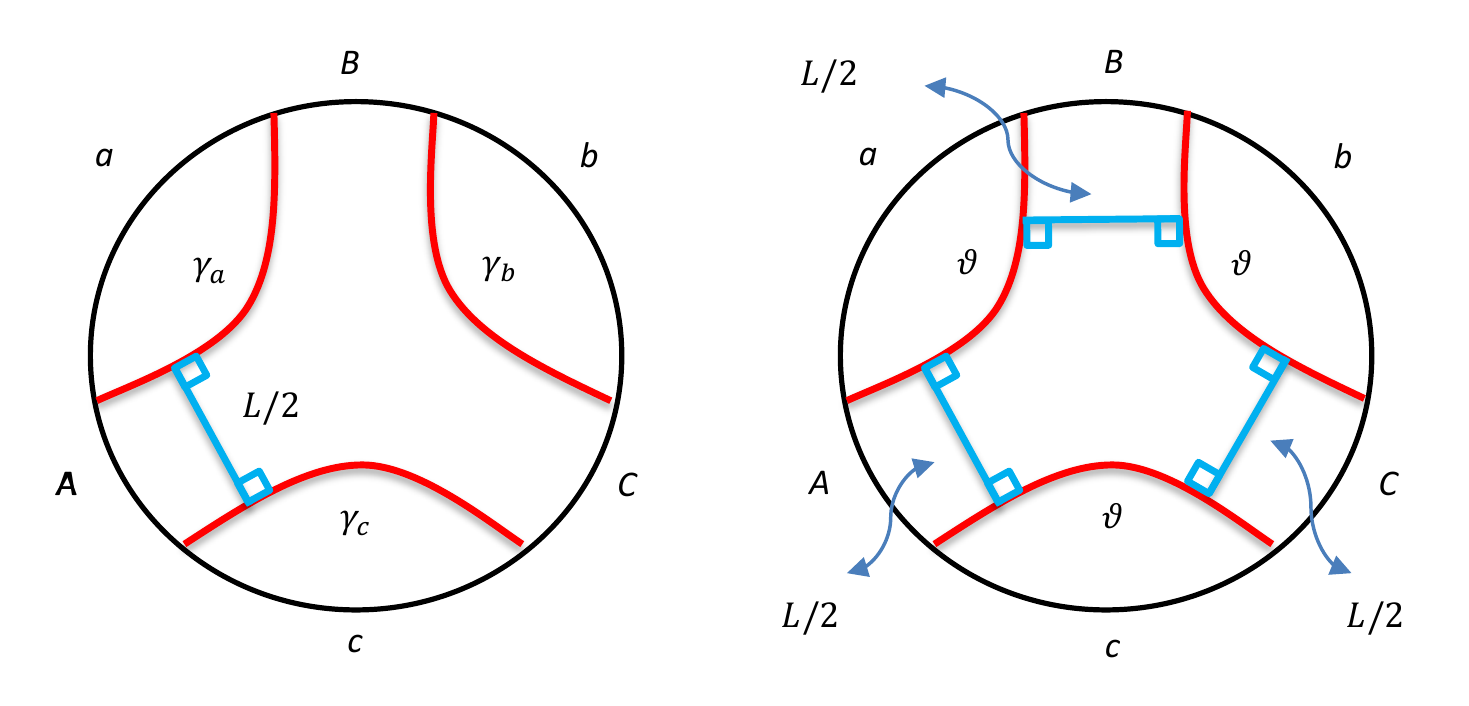}
\par\end{centering}
\caption{\label{fig:hexagon}The constructed   right-angled hexagon is bounded by the blue   and red geodesics.}
\end{figure}

\noindent The first step is to prepare a right-angled hexagon with
side lengths $L/2$, $\vartheta$, $L/2$ , $\vartheta$, $L/2$
, $\vartheta$ on a hyperbolic surface. Refer to Fig. (\ref{fig:hexagon}), the construction is
to begin with a Poincare disk and plot three boundary-anchored equal length
geodesics (in red color) $\gamma_{a}=\gamma_{b}=\gamma_{c}$ ( $a=b=c$ and $A=B=C$
on the boundary). Then, the hexagon is uniquely fixed by the following theorem:

\vspace*{2.0ex}

\noindent \textbf{Theorem} \textbf{1} (Ultra-parallel theorem, \cite{Buser}
Theorem 1.1.6): In the Poincare disk, if $\gamma_a$ and $\gamma_c$ are disjoint geodesics
with positive distance, then there exists a unique (blue colored) geodesic  perpendicular
to them.

\vspace*{2.0ex}

\noindent The lengths of the blue geodesics are denoted as $L/2$, which are of course the shortest lines
connecting $\gamma_i$.
The lengths of the other three red geodesics of  the   hexagon are  $\vartheta$. Trigonometry of
hyperbolic geometry \cite{Buser} (Theorem 2.4.1) gives
\begin{equation}
\cosh\vartheta =\frac{\cosh (L/2)}{\cosh (L/2)-1}
\label{eq: boundary length}
\end{equation}

\noindent Therefore, the   hexagon is enclosed by
the blue and red geodesics.

\subsubsection*{\uline{Step two:}}

\begin{figure}[H]
\begin{centering}
\includegraphics[scale=0.7]{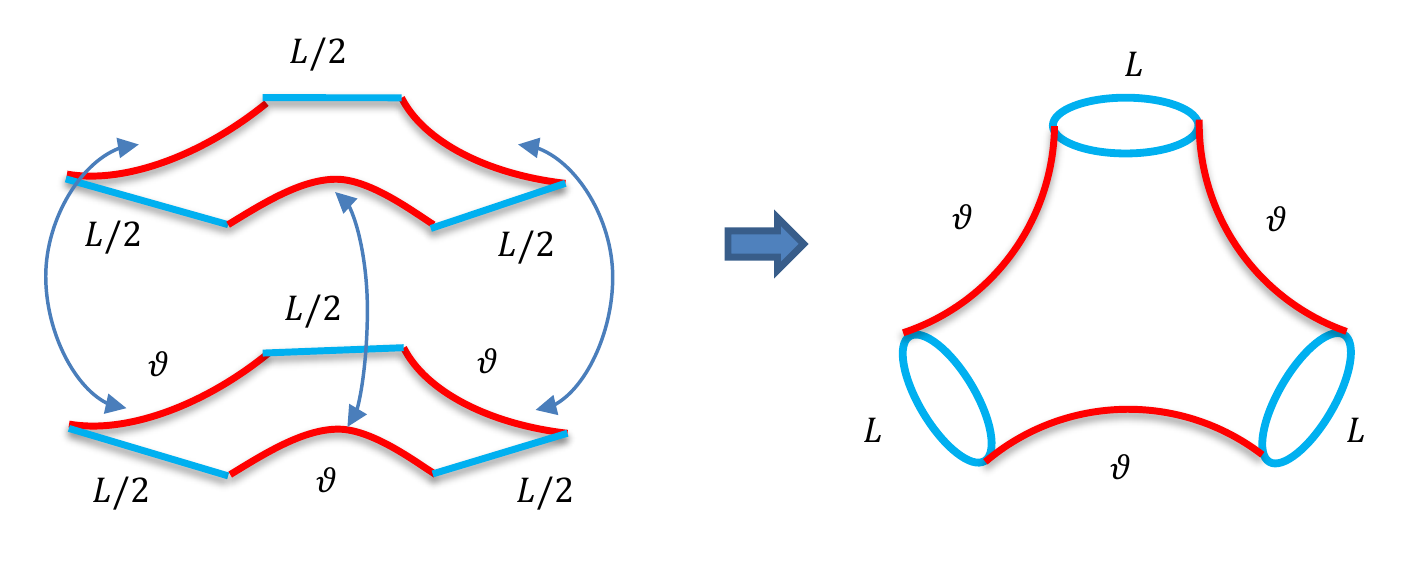}
\par\end{centering}
\caption{\label{fig:Y-pants CSFT}Glue two right-angled hexagons to obtain
a Y-piece.}
\end{figure}

Gluing two copies of this hexagon along the red geodesics $\vartheta$, we get an
$\tilde{\mathcal{V}}_{0,3}\left(L\right)$, called a
Y-piece whose three geodesic boundaries having lengths $\left(L/2\right)\times2=L$,
as depicted in Fig. (\ref{fig:Y-pants CSFT}).

\noindent The systole $sys[\Sigma]$ of a surface $\Sigma$ is defined as the length of
the shortest non-contractible closed geodesic
which is not a boundary component.  The construction itself guarantees
$sys[\tilde{\mathcal{V}}_{0,3}\left(L\right)] \ge L$.

\subsubsection*{\uline{Step three:}}

\begin{figure}[H]
\begin{centering}
\includegraphics[scale=0.7]{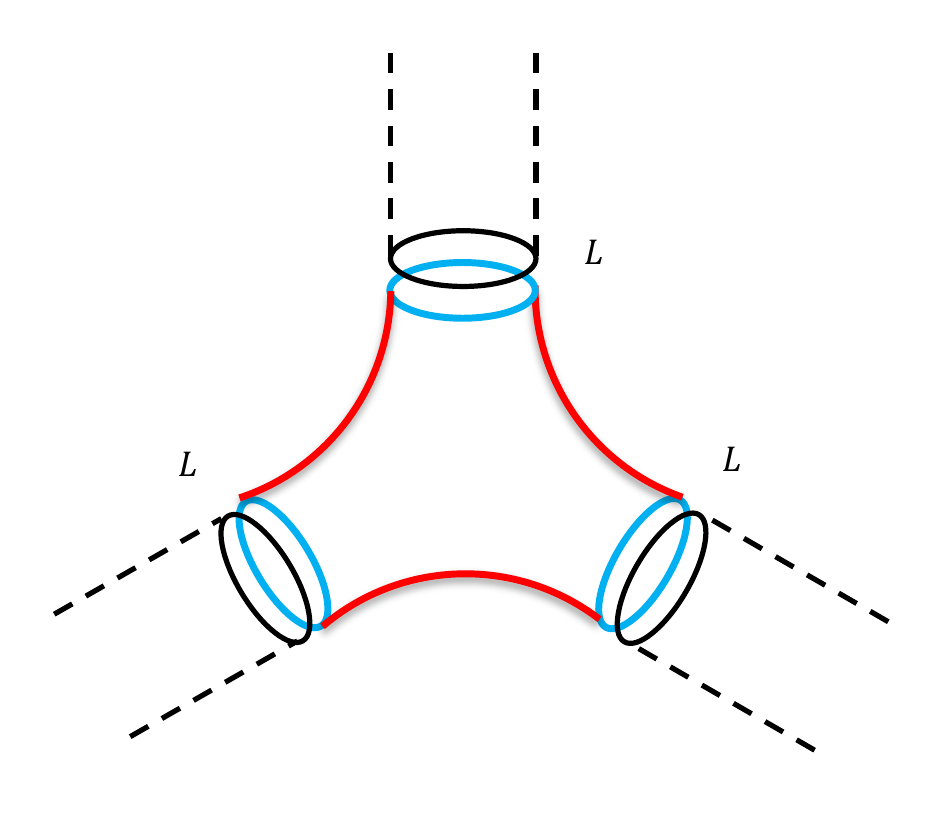}
\par\end{centering}
\caption{\label{fig:graft}$\tilde{\mathcal{V}}_{0,3}\left(L\right)\to
\mathcal{V}_{0,3}\left(L\right)$ by grafting semi-infinite cylinders.}
\end{figure}

The third step is to graft flat  semi-infinite  cylinders of circumference
$L$ to the  geodesic boundaries of $\tilde{\mathcal{V}}_{0,3}\left(L\right)$,
to obtain   the string vertices $\mathcal{V}_{0,3}\left(L\right)$, see Fig. (\ref{fig:graft}).

\noindent The flat semi-infinite cylinders  can be conformally mapped
to the marked points  with local coordinates.
Moreover, the cylinders can be seen as the Feynman propagators which
are used to connect different string vertices.
Higher order string vertices $\mathcal{V}_{g,n} (L)$ are
built by gluing   Y-piece $\tilde{\mathcal{V}}_{0,3}\left(L\right)$  along the
geodesic boundaries to form $\tilde{\mathcal{V}}_{g,n}\left(L\right)$, and
then grafting flat semi-infinite cylinders to the boundaries of $\tilde{\mathcal{V}}_{g,n}\left(L\right)$.

\subsubsection*{\uline{Step four}}

To solve
the geometric master equation (\ref{eq:BV equation}), the boundary geodesic length $L$ of  $\mathcal{V}_{g,n}(L)$ is
required to satisfy some constraints based on the Collar theorem.

\begin{figure}[H]
\begin{centering}
\includegraphics[scale=0.7]{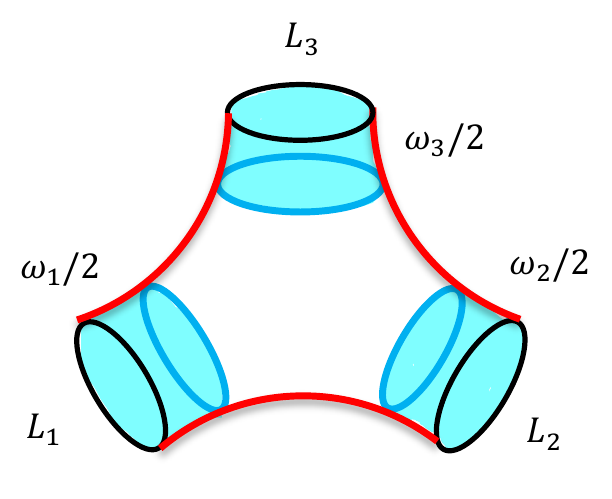}
\par\end{centering}
\caption{\label{fig:collar} Collars respecting eq. (\ref{eq: collar cond})  are pairwise disjoint.}
\end{figure}

\noindent \textbf{Theorem} \textbf{2} (Collar theorem, \cite{Buser} Theorem 4.1.1): Let $\sigma_i$ be  simple
closed geodesics on a hyperbolic surface $S$, the collars
\begin{equation}
\mathcal{C}\left(\sigma_i\right)=\left\{ p\in S|d\left(p,\sigma_i\right)\leq\frac{\omega_i}{2}\right\},
\end{equation}
of widths $\omega_i$
\begin{equation}
\sinh\left(\frac{1}{2}\omega_i\right)\sinh\left(\frac{1}{2}L\left(\sigma_{i}\right)\right)=1,
\label{eq: collar cond}
\end{equation}
are pairwise disjoint. We illustrate the simplest example $\tilde{\mathcal{V}}_{0,3}(L)$ in Fig. (\ref{fig:collar}).

\noindent  A critical geodesic length $L_*$ is defined as
\begin{equation}
\omega_{*}=L_{*}=2\mathrm{arc}\sinh\left(1\right)=2\log\left(1+\sqrt{2}\right)=\log\left(3+2\sqrt{2}\right).
\end{equation}
A very useful proposition can be proved with the Collar theorem:

\vspace*{1.0ex}

\noindent \textbf{Proposition 3} (\cite{Buser} Proposition 4.1.2): Let $\sigma$, $\delta$ be closed
geodesics on a hyperbolic surface which intersect each other transversally, and assume $\sigma$ is simple. Then
\begin{equation}
\sinh\left(\frac{\sigma}{2} \right) \sinh\left(\frac{\delta}{2} \right) >1.
\end{equation}
An immediate conclusion is that a simple geodesic of length $L\le L_*$ cannot intersect
another geodesic of length $L' \le L_*$.
Based on these conclusions,
in \cite{Costello:2019fuh},  Costello and  Zwiebach proved:
\begin{enumerate}
\item The sets $\mathcal{V}\left(L\right) = \sum_{n\ge 3} \mathcal{V}_{0,n}\left(L\right)$
whose  boundary length is $L>0$ and $sys[\mathcal{V}(L)]\ge L$,
solve the classical geometric master equation ($\hbar\rightarrow0$
in (\ref{eq:BV equation})).
\item The sets $\mathcal{V}\left(L\right) = \sum_{g,n} \mathcal{V}_{g,n}\left(L\right)$
whose  boundary length is $L\le L_* $ and $sys[\mathcal{V}(L)]\ge L$,
solve the quantum geometric master equation (\ref{eq:BV equation}).
\end{enumerate}
Note that as Costello and Zwiebach demonstrated, for $0< L\le L_*$, the operation $\Delta$
and $\{,\}$ in the BV equation (\ref{eq:BV equation}) maintain the systole
of $\tilde{\mathcal{V}}_{g,n}\left(L\right)$. When constructing $\tilde{\mathcal{V}}_{g,n}\left(L\right)$ from $\tilde{\mathcal{V}}_{0,3}\left(L\le L_*\right)$,   the systole does not change
according to Proposition $3$.

%

\subsection{Canonical purification and reflected entropy}

In this subsection, we provide a brief introduction to the
reflected entropy and its bulk dual, namely reflected surfaces.
Some recent progresses refer to \cite{Bao:2021vyq}.

Our discussion starts
with the entanglement entropy of a pure state in CFT$_{2}$. Considering
a quantum pure system $|\psi\rangle$  which is divided into two parts $A$ and $B$, the total Hilbert space is $\mathcal{H}=\mathcal{H}_{A}\otimes\mathcal{H}_{B}$. The reduced density matrix of the subsystem $A$ is defined
as $\rho_A = \mathrm{Tr}_{\mathcal{H}_B}|\psi\rangle\langle \psi|$. The entanglement
entropy of  the  region $A$ is given
by the von Neumann entropy:

\begin{equation}
S_{EE}\left(A\right)=-\mathrm{Tr}_{\mathcal{H}_{A}}\left(\rho_{A}\log\rho_{A}\right).
\end{equation}
It is  clear that $S_{EE}\left(A\right)=S_{EE}\left(B\right)$.

\begin{figure}[H]
\begin{centering}
\includegraphics[scale=0.7]{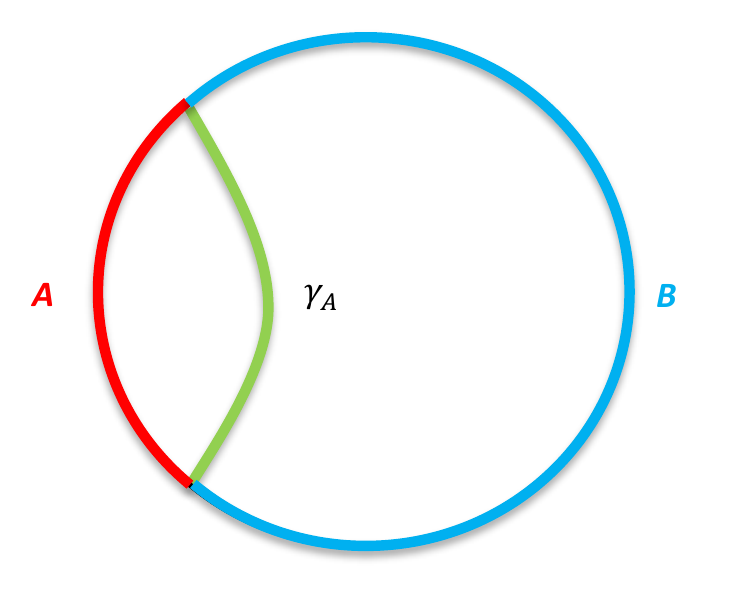}
\par\end{centering}
\caption{\label{fig:RT} The entanglement entropy $S_{EE}\left(A\right)$ can
be evaluated by the length of geodesic $\gamma_{A}$ (green line).}
\end{figure}

Ryu and Takayanagi (RT)  \cite{Ryu:2006bv} realized that this entanglement
entropy possesses a bulk interpretation. As illustrated in Fig. (\ref{fig:RT}) for
AdS$_3$/CFT$_2$, the bulk dual of the entanglement entropy is the area of the
codimension-$2$ minimal surface $\gamma_A$ (green line, geodesics for $d=2$)
satisfying: $\partial A = \partial \gamma_A$
and $\gamma_A$ is homologous to $A$:

\begin{equation}
S_{EE}\left(A\right)=\frac{Area\left(\gamma_{A}\right)}{4G_{N}^{(3)}},
\end{equation}

\noindent where $G_{N}^{(3)}$ is the $3$-dimensional Newton constant.
We will set $4G_{N}^{(3)}=1$ for simplicity in the following discussions.

However, the entanglement entropy fails if the concerned state $\rho_{AB}$ is a mixed state.
The concept of entanglement of purification  is thus introduced  \cite{Terhal} to measure the entanglement
of mixed states. We focus on bipartite mixed states here.

Suppose there is a
mixed state $\rho_{AB} = \sum_i\, p_i |i\rangle_{AB} \langle i|_{AB}$, organized in terms of an orthonormal basis.
By enlarging the Hilbert space from $\mathcal H_A\otimes \mathcal H_B$
to $\mathcal H_A\otimes \mathcal H_B \otimes \mathcal H'_A\otimes \mathcal H'_B$, the mixed
state $\rho_{AB}$ is purified to a pure state $\left| \sqrt{\rho_{AB}}  \right\rangle _{AA^{\prime}
BB^{\prime}}= \sum_i \, \sqrt{p_i} |i\rangle_{AB} \otimes |i\rangle_{A'B'} $.
It should be noted that the purification is far from unique.
Then the entanglement of purification
is defined as the von Neumann entropy between
$A\cup A^{\prime}$ and $B\cup B^{\prime}$:  $E_{P}\left(\rho_{AB}\right)
=\underset{\left|\psi\right\rangle _{AA^{\prime}BB^{\prime}}}{\mathrm{min}}S_{EE}\left(A\cup A^{\prime}\right)$,
minimizing over all possible purifications.

\begin{figure}[H]
\begin{centering}
\includegraphics[scale=0.7]{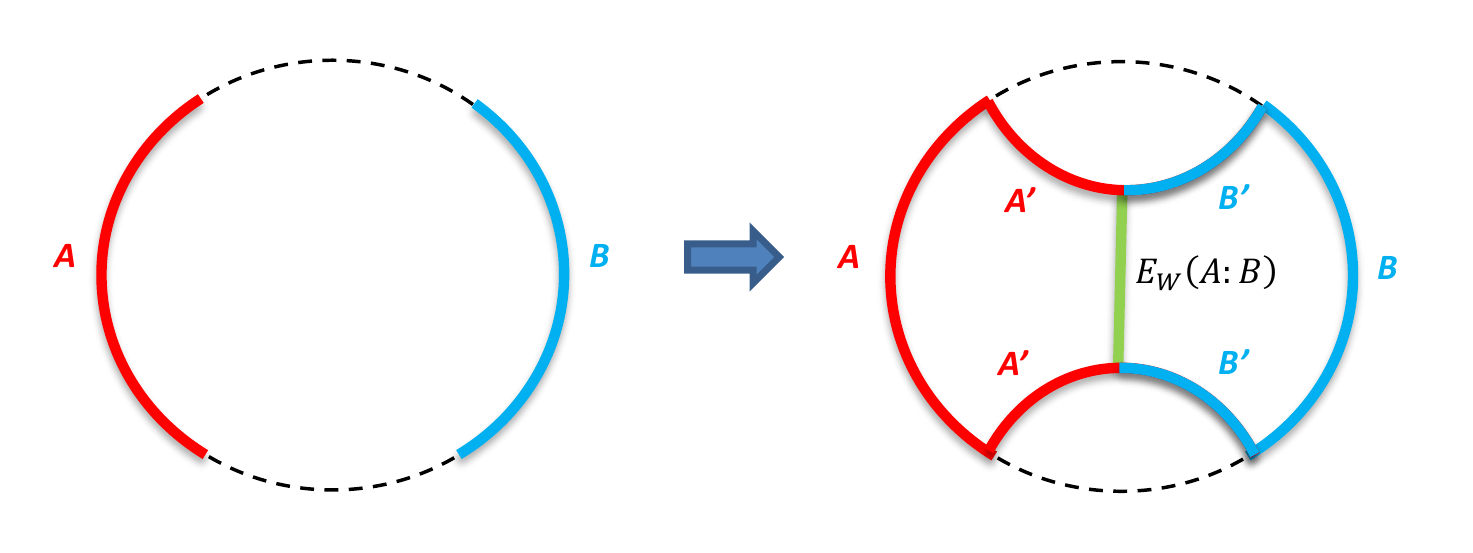}
\par\end{centering}
\caption{\label{fig:puri} Purification of mixed states. The green line is
the entanglement wedge cross-section $E_{W}$.}
\end{figure}

The bulk interpretation of $E_P$ is shown in Fig. (\ref{fig:puri}). In this
figure,  the subsystems $A$ and $B$ are
no longer complementary on the boundary. The procedure of purification
provides the bulk geodesic regions $A^{\prime}$ and $B^{\prime}$.
Then the new subsystems $A\cup A^{\prime}$ and $B\cup B^{\prime}$
become complementary again.
The entanglement wedge cross-section (EWCS)
$E_{W}$,  the green line in Fig. (\ref{fig:puri}), is defined by minimizing
the RT surface of $A\cup A^{\prime}$ and $B\cup B^{\prime}$ over $A^{\prime}$ and $B^{\prime}$.
It is conjectured that \cite{Takayanagi:2017knl}:

\begin{equation}
E_{P}\left(\rho_{AB}\right) =E_{W}\left(A:B\right).\label{eq:EW}
\end{equation}

\noindent This  is a generalization of the pure state situation. If $\rho_{AB}$ is
a pure state, one has  $E_P (\rho_{AB})  =E_{W}\left(A:B\right)=S_{EE}\left(A\right)=S_{EE}\left(B\right)$.

However,  the entanglement of purification $E_{P}\left(\rho_{AB}\right)$
is difficult to calculate since it involves minimization over all possible purifications.
For this reason,  in \cite{Dutta:2019gen}, a much simpler alternative has been proposed,
namely the reflected entropy $S_{R}$.

Instead of minimizing over all possible purifications, the reflected entropy simply selects an identical
copy of the original mixed state as the purification, which is called canonical purification. Denoting the
copied state as $A^* B^*$, the purified state and the reflected entropy are

\begin{equation}
\left| \sqrt{\rho_{AB}}  \right\rangle _{AA^*
BB^*}= \sum_i \, \sqrt{p_i} |i\rangle_{AB} \otimes |i\rangle_{A^* B^*},\quad\quad S_{R}\left(A:B\right)=S\left(AA^{*}\right)_{\sqrt{\rho_{AB}}}= -{\rm Tr} \rho_{AA^*} \log \rho_{AA^*},
\end{equation}
where $\rho_{AA^*} = {\rm Tr}_{BB^*}  | \sqrt{\rho_{AB}}  \rangle  \langle \sqrt{\rho_{AB}} |$.
If $A\cup B$ is a pure state, the reflected entropy
$S_{R}\left(A:B\right)$ reduces to a double of the ordinary entanglement entropy
$S_{R}\left(A:B\right)=2S_{EE}\left(A\right)=2S_{EE}\left(B\right)$.

\begin{figure}[H]
\begin{centering}
\includegraphics[scale=0.7]{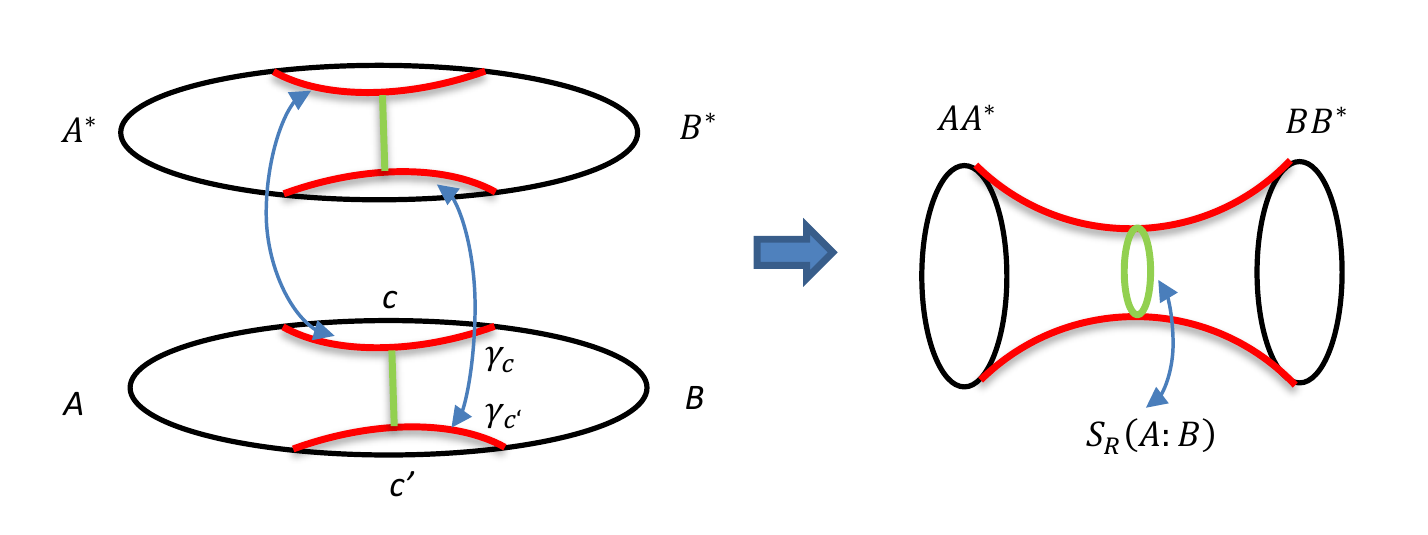}
\par\end{centering}
\caption{\label{fig:puri example} The surface $\left(A^{*},B^{*},c,c^{\prime}\right)$ is
a CPT copy of the original surface $\left(A,B,c,c^{\prime}\right)$. $\gamma_{c}$
and $\gamma_{c^{\prime}}$ are geodesics
anchored on $\partial c$ and $\partial c'$ respectively. The reflected surface $S_R(A:B)$ is
a closed simple geodesic, obtained by gluing   along $\gamma_{c}$ and $\gamma_{c^{\prime}}$.}
\end{figure}

The bulk dual of the  reflected entropy is called reflected surface,  which is also denoted as $S_R(A:B)$.
For a bipartite mixed state  $\left(A,B,c,c^{\prime}\right)$, as illustrated  in Fig. (\ref{fig:puri example}),
we first prepare  a CPT copy  $\left(A^{*},B^{*},c,c^{\prime}\right)$.
The red curves $\gamma_{c}$ and $\gamma_{c^{\prime}}$ are geodesics anchored on $\partial c$ and $\partial c'$ respectively.
The reflected surface $S_R(A:B)$ (green curve) is obtained by gluing two surfaces along
$\gamma_{c}$ and $\gamma_{c^{\prime}}$. By definition, the reflected surface is a \emph{closed simple geodesic} \cite{Dutta:2019gen,Engelhardt:2018kcs} and  one thus has

\begin{equation}
S_{R}\left(A:B\right)=2E_{W}\left(A:B\right),
\end{equation}

\begin{figure}[H]
\begin{centering}
\includegraphics[scale=0.7]{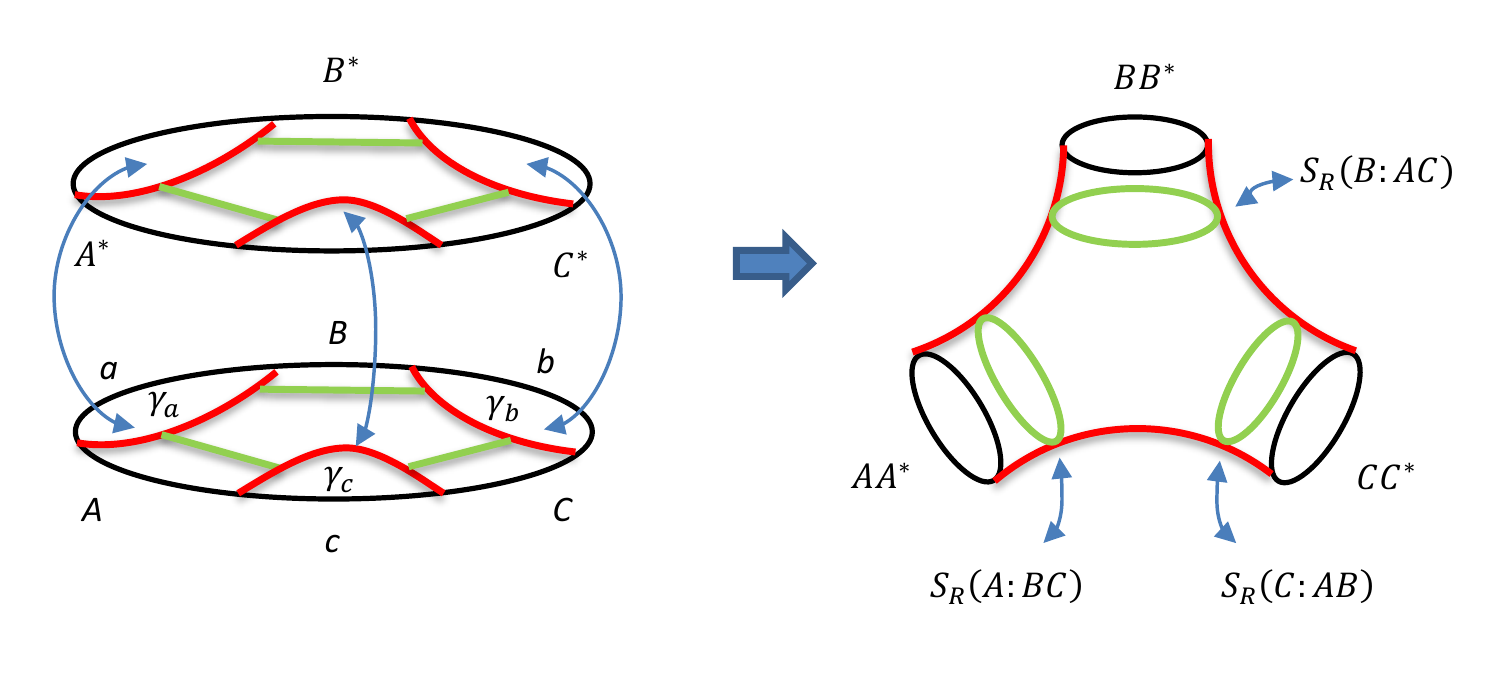}
\par\end{centering}
\caption{\label{fig:tri puri}Bulk interpretation for tripartite canonical
purification.}
\end{figure}

\noindent These constructions can be   generalized to multipartite mixed
states \cite{Bao:2019zqc, Chu:2019etd}.
The canonical purification of a tripartite mixed state is dual to gluing
two copies of tripartitioned entanglement wedge along three red geodesics
in the bulk, see Fig. (\ref{fig:tri puri}).

\noindent The lengths of green simple closed geodesics (reflected
surfaces) are reflected entropies for bipartitions $\left(A:BC\right)$,
$\left(B:AC\right)$ and $\left(C:AB\right)$ in the boundary theory\cite{Chu:2019etd}:

\begin{eqnarray}
S_{R}\left(A:BC\right) & = & S\left(AA^{*}:BB^{*}CC^{*}\right)_{\sqrt{\rho_{ABC}}},\nonumber \\
S_{R}\left(B:AC\right) & = & S\left(BB^{*}:AA^{*}CC^{*}\right)_{\sqrt{\rho_{ABC}}},\nonumber \\
S_{R}\left(C:AB\right) & = & S\left(CC^{*}:AA^{*}BB^{*}\right)_{\sqrt{\rho_{ABC}}}.
\end{eqnarray}

\subsubsection*{\uline{Lower bound of $S_{R}$}}

\begin{figure}[H]
\begin{centering}
\includegraphics[scale=0.7]{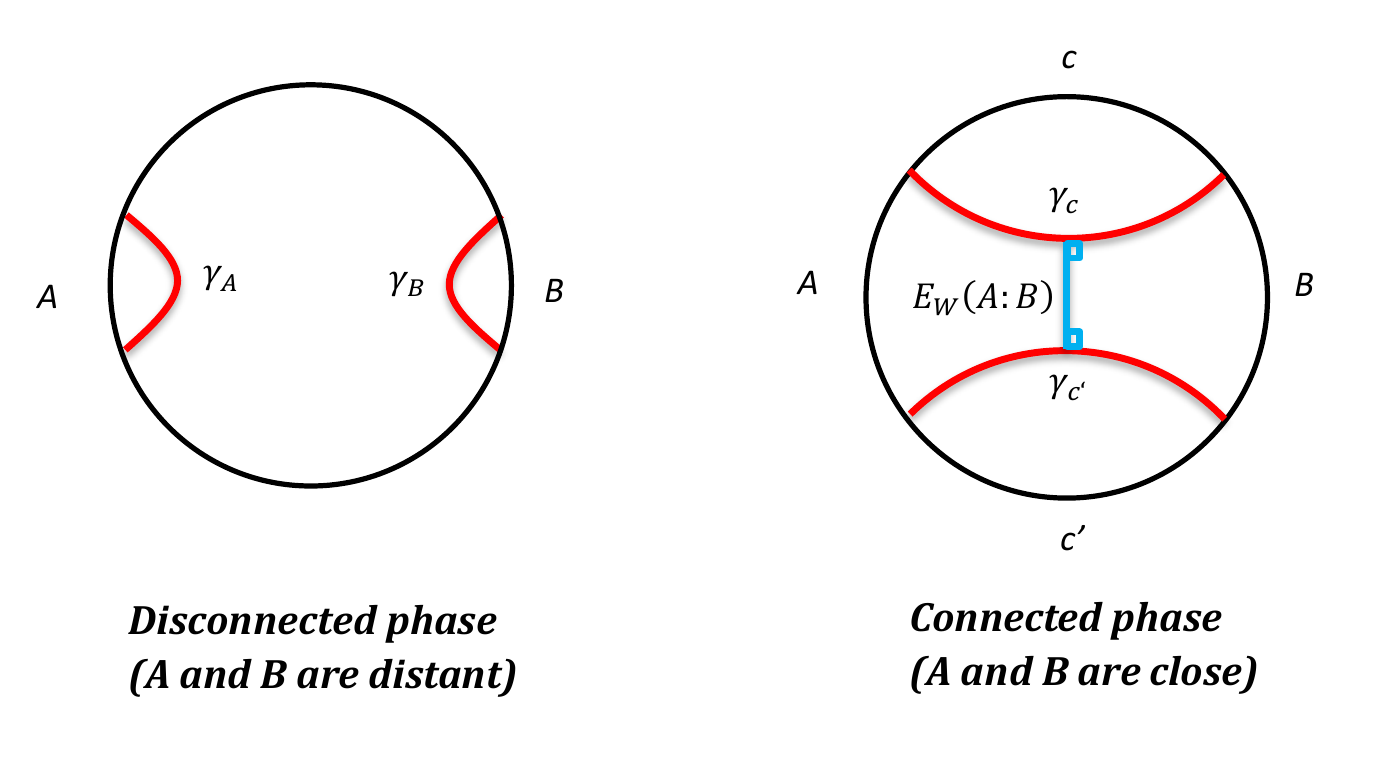}
\par\end{centering}
\caption{\label{fig:phase}Phase transition between disconnected and connected
phases.}
\end{figure}

Unlike the boundary of hyperbolic string vertices, whose lengths have an upper limit $L\le L_*$,
the length of a reflected surface $S_{R}$  has a lower bound. This bound
comes from the definition of the entanglement wedge cross section:

\begin{equation}
E_{W}\left(A:B\right)\geq\frac{1}{2}I\left(A:B\right),
\end{equation}

\noindent where $I\left(A:B\right)=S\left(\rho_{A}\right)+S\left(\rho_{B}\right)-S\left(\rho_{AB}\right)$
is the mutual information. When $A$ and $B$ are distant, the mutual
information $I\left(A:B\right)$ vanishes \cite{Headrick:2010zt},
and therefore the corresponding bulk entanglement wedge becomes disconnected,
namely $E_{W}\left(A:B\right)=0$, see Fig. (\ref{fig:phase}). In
other words, the reflected entropy or entanglement wedge cross section
does not always exist. There is a phase transition point between the
connected and disconnected entanglement wedges.

\noindent The lower bound of $S_{R}\left(A:B\right)$ and $I\left(A:B\right)$
can be obtained either from CFT or geodesics in AdS  (below
eq. (21) in ref. \cite{Takayanagi:2017knl} and below eq. (4.40) in
ref. \cite{Dutta:2019gen}). We simply present their results here.
Suppose there are two regions $A=\left[a_{1},a_{2}\right]$ and $B=\left[b_{1},b_{2}\right]$,
where $a_{1}<a_{2}<b_{1}<b_{2}$, the entanglement wedge cross section
and the mutual information are given by

\begin{eqnarray}
E_{W}\left(A:B\right) & = & \log\left(1+2z+2\sqrt{z\left(z+1\right)}\right),\nonumber \\
\frac{1}{2}I\left(A:B\right) & = & \log z,\label{eq:EW and mutual}
\end{eqnarray}

\noindent where we set $R=4G_{N}=1$ and $z\geq0$ is the cross ratio,

\begin{equation}
z=\frac{\left(a_{2}-a_{1}\right)\left(b_{2}-b_{1}\right)}{\left(b_{1}-a_{2}\right)\left(b_{2}-a_{1}\right)}.
\end{equation}

\noindent From eq. (\ref{eq:EW and mutual}), $I\left(A:B\right)$ is truncated at $z=1$. Therefore, at this value, $E_{W}\left(A:B\right)$ reaches a lower bound,

\begin{equation}
E_{W}\left(A:B\right)>\log\left(3+2\sqrt{2}\right).
\end{equation}

\noindent Then, it is ready to get

\begin{equation}
S_{R}\left(A:B\right)=2E_{W}\left(A:B\right)>2\log\left(3+2\sqrt{2}\right)=2L_{*},
\end{equation}

\noindent which indicates that when the length of the green geodesic $S_{R}\left(A:B\right)$
approaches $2L_{*}$ in Fig. (\ref{fig:puri example}),
the entanglement wedge, which is enclosed by the red geodesics, abruptly disappears
and $S_{R}\left(A:B\right)$ becomes zero. In addition, the tripartite
entanglement wedge also requires the three bipartite reflected entropies satisfying
$S_{R}\left(A:BC\right)$, $S_{R}\left(B:AC\right)$ and $S_{R}\left(C:AB\right)\geq2L_{*}$.

\section{Connections  between string vertices and reflected entropies}

We now see that both the boundaries of string vertices and reflected surfaces are simple closed geodesics
in hyperbolic surfaces.  According to Theorem $1$, there is a unique geodesic perpendicular to two
ultra-parallel geodesics. We also have:
%
%
%
%

\vspace*{1.0ex}

\noindent \textbf{Theorem 4} (\cite{Buser} Theorem 3.1.7): Given any three positive numbers
$l_{a}$, $l_{b}$, $l_{c}$, there exists a unique Y-piece whose boundary geodesics have
lengths $l_{a}$, $l_{b}$, $l_{c}$.

\vspace*{1.0ex}

\noindent Furthermore, the Y-pieces are the fundamental
building blocks for higher order string vertices and reflected surfaces.
So, it is reasonable to expect there are some close connections between them.

\begin{figure}[H]
\begin{centering}
\includegraphics[scale=0.7]{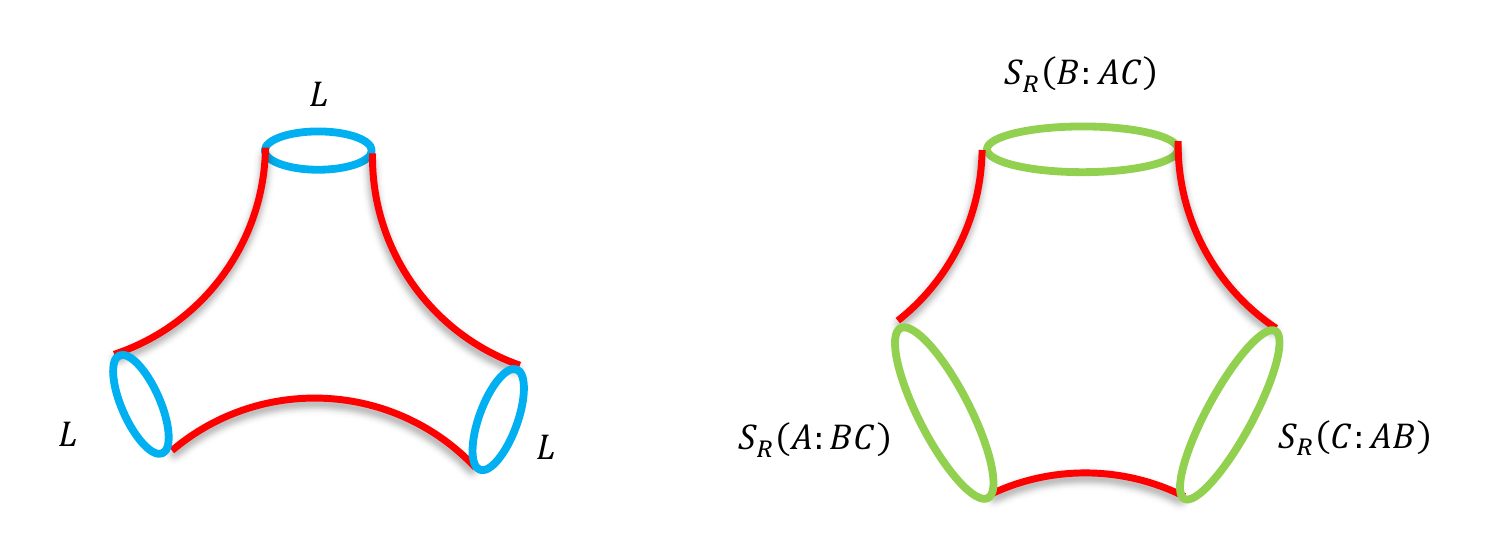}
\par\end{centering}
\noindent \begin{centering}
\begin{tabular}{|c|c|c|}
\hline
 & Closed string vertices $\tilde{\mathcal{V}}_{0,3}\left({\color{blue}L}\right)$ & Reflected entropies ${\color{green}S_{R}\left(\bullet:\bullet\bullet\right)}$\tabularnewline
\hline
Constraint 1 & Equal lengths of boundaries & No requirement\tabularnewline
\hline
Constraint 2 & ${\color{blue}L}\leq L_{*}$ for quantum vertices & ${\color{green}S_{R}\left(\bullet:\bullet\bullet\right)}>2L_{*}$\tabularnewline
 & ${\color{blue}L}>0$ for classical vertices & \tabularnewline
\hline
\end{tabular}
\par\end{centering}
\caption{\label{fig:comparation}Differences of Y-pieces in two theories.}
\end{figure}

However, the connection is not   straightforward. As discussed in last section,
the boundary lengths of the string vertices are constrained to be the same while
those of reflected surfaces are allowed to be different. A more serious problem is that
the  boundary lengths of the (quantum) string vertices have an upper bound $L\le L_*$, but
those of reflected surfaces have a lower bound $S_R > 2L_*$. We summarize these differences in
 Fig. (\ref{fig:comparation}) and the table,

%
%
%
%
%


One might be puzzled why we introduce the string vertices to complicate the story.
It looks that the Y-pieces and surfaces built on them are sufficient to construct
the reflected surfaces, without the two strict constraints imposed by the string vertices.
The reasons are three folds:
\begin{enumerate}
\item Though the classical string vertices allow any boundary geodesic length, they are
only applicable to tree diagrams. To see this, let us recall the BV equation

\begin{equation*}
\partial\mathcal{V}+\hbar\triangle\mathcal{V}+\frac{1}{2}\left\{ \mathcal{V},\mathcal{V}\right\} =0.
\end{equation*}

\noindent In the classical limit $\hbar\to 0$, the second term $\hbar\triangle\mathcal{V}$ disappears,
and  $\mathcal{V}=\underset{g,n}{\sum}\hbar^{g}\mathcal{V}_{g,n}\to \mathcal{V}=\underset{n\ge 3}{\sum}\mathcal{V}_{0,n}$.
So, we will mainly discuss the quantum vertices and mention the classical limits when needed.

For the  string vertices,  BV equation $=$ upper bounded hyperbolic surfaces:  $L\le L*$.
On the other hand,  the reflected surfaces are lower bounded hyperbolic surfaces: $S_{R}>2L_{*}$,
constrained from the mutual information.
Both objects are thus subsets of hyperbolic surfaces. Therefore, building connections
between them is interesting and informative.
One can anticipate some nontrivial results emerged from the connections.

\item The string vertices  respect the  dynamical equation, namely
the BV equation (\ref{eq:BV equation}).  So, the string vertices provide us a practical approach to
figure out dynamical properties of the reflected surfaces and   entropies.

\item Through the connection, it is possible to build in CFT the holographic duals of
the operations $\Delta$ and $\{,\}$.
\end{enumerate}
We first show how to construct reflected surfaces from string vertices,
and then address the  equal length restriction of the string vertices.

\subsection{Construct  reflected surfaces from string vertices}

In this subsection,  we present three examples to show how
to construct the bulk geometries of multipartite canonical purification
from the string vertices $\mathcal{V}_{0,3}\left(L\right)$. \emph{Since a hyperbolic Riemann surface has
different  decomposition patterns, our  constructions
are far from unique}.

%
%

\vspace*{1.0ex}
\uline{Example $1$}
\vspace*{1.0ex}

\begin{figure}[H]
\begin{centering}
\includegraphics[scale=0.7]{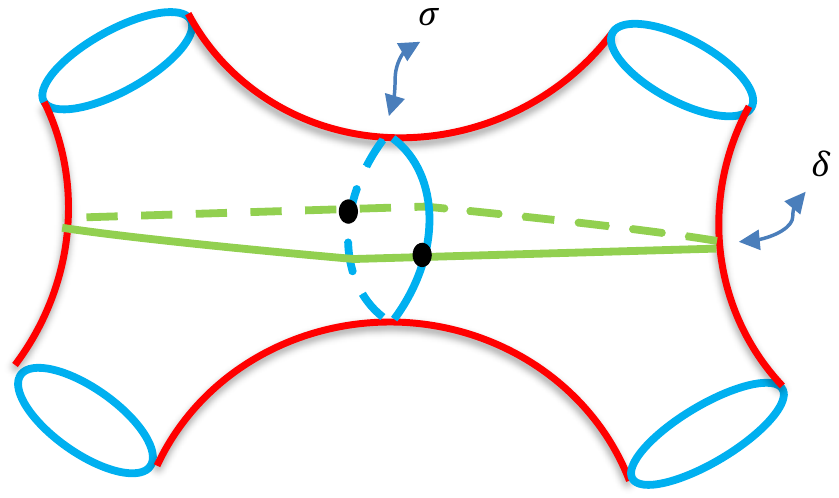}
\par\end{centering}
\caption{\label{fig:X-piece}Gluing two string Y-pants to get X-piece.}
\end{figure}

Let us start with  Fig. (\ref{fig:X-piece}), the X-piece
$\tilde{\mathcal{V}}_{0,4}\left(L\right)$, constructed by gluing two
Y-pieces along the blue geodesic boundary $\sigma$.

\noindent  According to Theorem $4$, the X-piece is uniquely fixed by the blue geodesics,
and we have:

\vspace*{1.0ex}

\noindent \textbf{Corollary} \textbf{5}: Considering two
closed geodesics with lengths $\sigma$ and $\delta$ on a X-piece\footnote{We use the same
notations for the lengths of the curves},
if $\sigma$ and $\delta$ intersect twice,  we have $max\left\{\sigma,\delta\right\} >2L_{*}$.
This bound is sharp \cite{Gauglhofer}.

\vspace*{1.0ex}

\noindent This result naturally combines the bounds of two theories together: $L\le L_*$ in
string vertices and $S_R > 2L_*$ in entanglement, and it makes it possible to establish
connections between two theories. More specifically, the quantitative relation between
the values of $\sigma$ and $\delta$ is (\cite{Buser}, Theorem 2.3.4 (i)):

\begin{equation}
\cosh\frac{\sigma }{2}=\sinh\frac{\delta}{4}\sinh\frac{\sigma}{4}.
\label{eq: Colollary 5}
\end{equation}


\noindent It is easy to derive that $\delta$ has a minimum,

\begin{equation}
\delta = 4 \mathrm{arc}\sinh \left(  \frac{\cosh\sigma/2}{\sinh\sigma/4}  \right) \ge 4 \mathrm{arc}\sinh(2\sqrt 2) \equiv \delta_{min} >  S_{Rmin}\equiv  4 \mathrm{arc}\sinh(1) =2 L_* .
\label{eq: delta}
\end{equation}
Note there is  a \emph{nontrivial} gap between
the minimal values   $\delta_{min}$ and $S_{Rmin}$.
Therefore, whatever the length of the blue boundary geodesics is, the green geodesic
is always a valid reflected surface which indicates  an up-down entanglement\footnote{Of course, one can rotate
the X-piece by $90$ degree and switch the roles of $\sigma$ and $\delta$. In this case, we always
have a left-right  entanglement.}. This realization provides a better option for the bulk interpretations for the reflection entropies.

In the classical theory,   if the length of $\sigma$ is
also larger than $2 L_*$,   Fig. (\ref{fig:X-piece}) basically represents a  quadripartite reflected
surface with four $S_{R}\left(\bullet:\bullet\bullet\bullet\right) =\sigma$,
one  $S_{R}\left(\bullet\bullet:\bullet\bullet\right) =\sigma$ and
one  $S_{R}\left(\bullet\bullet:\bullet\bullet\right) =\delta$.

\begin{figure}[H]
\begin{centering}
\includegraphics[scale=0.7]{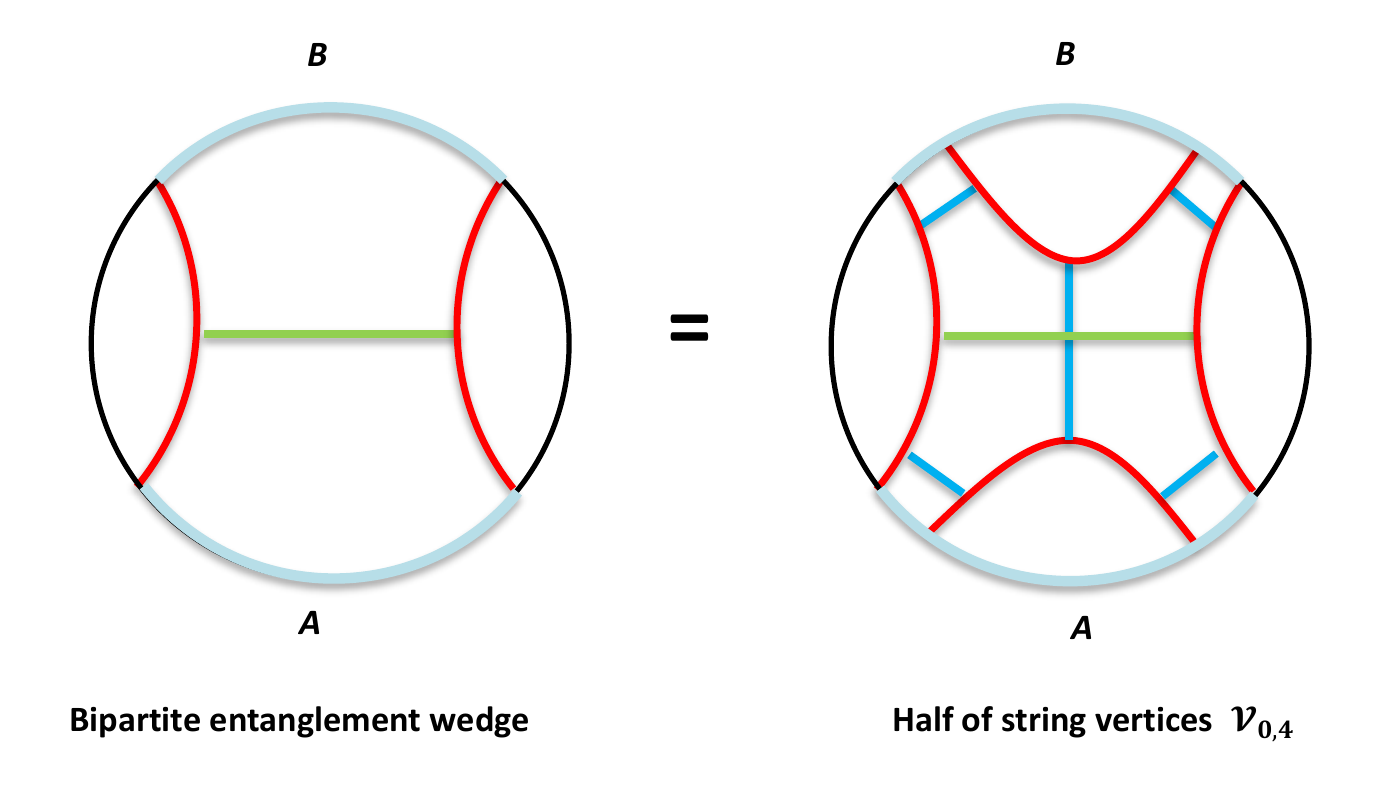}
\par\end{centering}
\caption{\label{fig:bi corres} A quantum X-piece represents a bipartite reflected surface.
Since the length of the  blue geodesics  is smaller than $L_*/2$, they
can not be reflected surfaces and no entanglement wedge exists. The only
legitimate reflected surface is the green geodesic which indicates an up-down entangling.}
\end{figure}

Note that in quantum theory, blue geodesics $\sigma < L_*$, so they do not contribute to the reflected entropy and the corresponding entanglement wedges disappear. We are therefore only left with the up-down bipartite  entanglement $S_{R}\left(\bullet:\bullet\right) =\delta$, namely the green geodesic,
as illustrated in    Fig. (\ref{fig:bi corres}), where the right panel collapses to the left panel.
From  eq. (\ref {eq: Colollary 5}),
a quantitative relation between the boundary length of the string vertices $L = \sigma$ and
the reflected entropy $S_R(A:B) = \delta$ is

\begin{eqnarray}
{\color{green}S_{R}}(A:B) & = & 4\mathrm{arc}\sinh\frac{\cosh\frac{{\color{blue}L}}{2}}{\sinh\frac{{\color{blue}L}}{4}},\label{eq:S}\\
{\color{blue}L} & = & 4\mathrm{arc}\sinh\left[\frac{1}{4}\left(\sinh\frac{{\color{green}S_{R}}}{4}\pm\sqrt{\sinh^{2}\frac{{\color{green}S_{R}}}{4}-8}\right)\right].\label{eq:L}
\end{eqnarray}

\vspace*{2.0ex}
\noindent\uline{The gap between $\delta_{min}$ and   $S_{Rmin}$, \emph{generalized} reflected entropy}
\vspace*{1.0ex}

In eq. (\ref{eq: delta}),
there exists a nontrivial gap between
the minimal values of $\delta_{min}=4 \mathrm{arc}\sinh(2\sqrt 2)$ and $S_{Rmin}= 2 L_*=4 \mathrm{arc}\sinh(1)$.
It is very instructive to study the origin of this gap.

\begin{figure}[H]
\begin{centering}
\includegraphics[scale=0.7]{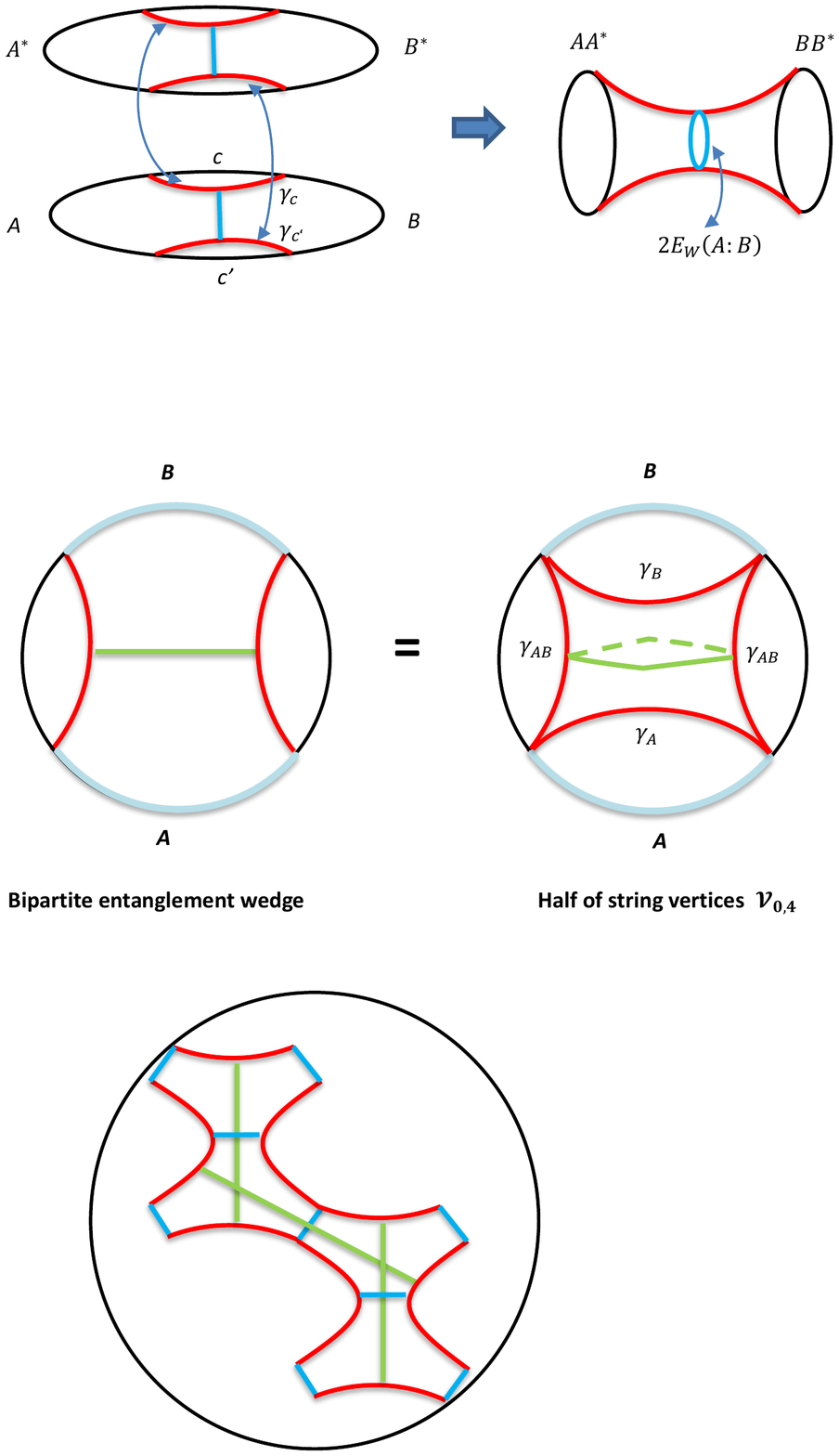}
\par\end{centering}
\caption{\label{fig:four string}Naive four-string vertex: it is a limit of Costello-Zwiebach hyperbolic string vertex.}
\end{figure}

The reason is that the hyperbolic string vertices with   explicit
local coordinates provide   non-perturbative corrections to the reflected entropies.
In \cite{Firat:2021ukc}, it shows that the hyperbolic three-string vertex reduces to the naive three-string
vertex as the geodesic boundaries $L$'s approach zero. It is not difficult to generalize this result to the
four-string vertex.
Under this limit, the outer four geodesic boundaries $L$'s of $\mathcal{V}_{0,4}$ go to zero, one then gets $\delta_{min}\rightarrow S_{Rmin}= 2 L_*$ (Proposition
3.1. in ref. \cite{Gauglhofer:2007}), as depicted in Fig. (\ref{fig:four string}).
It therefore indicates that the real bulk dual  of the reflected entropy
is not the reflected surface (hyperbolic string vertex), but the Moosavian-Pius surface!
Since the Moosavian-Pius surface is a limit of the hyperbolic string vertex,
the hyperbolic string vertex should be dual to a \emph{generalized} reflected entropy in the CFT, which has
the reflected entropy as a limit.

This limiting behavior is also consistent with a bulk
interpretation of the mutual information:

\begin{eqnarray}
I\left(A:B\right) & = & S_{A}+S_{B}-S_{AB}\nonumber \\
 & = & Area\left(\gamma_{A}\right)+Area\left(\gamma_{B}\right)-Area\left(2\gamma_{AB}\right).
\end{eqnarray}

Therefore, the hyperbolic string vertices provide   alternative quantum corrections to the entanglement
entropy. Since the adjacent red geodesics cannot join together at one point on the boundary in Fig. (\ref{fig:bi corres}), the quantum gravitational effects (originated from closed strings) in the bulk will introduce a  minimal observable length near the boundary, and then modify the minimal value of the reflected entropy  $S_{Rmin}$ to $\delta_{min}$ in eq. (\ref{eq: delta}).

\begin{figure}[H]
\begin{centering}
\includegraphics[scale=0.6]{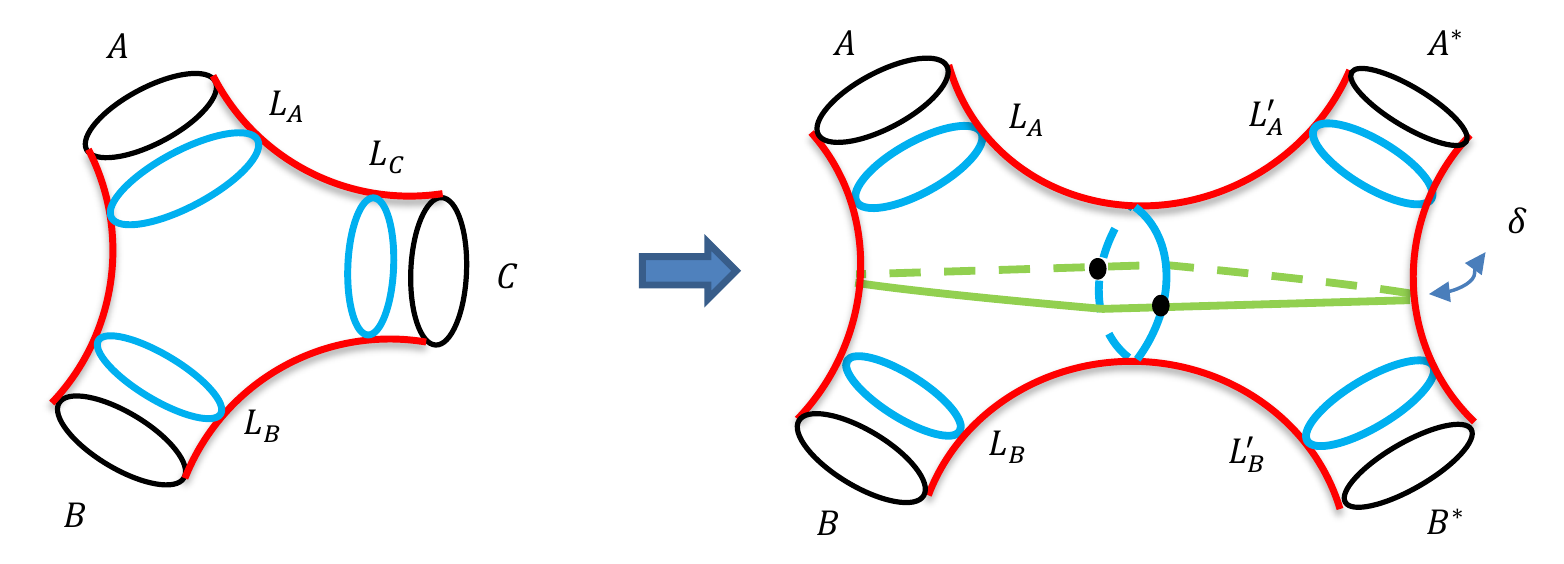}
\par\end{centering}
\caption{\label{fig:worm puri} The canonical purification of three boundary
wormhole.}
\end{figure}

This gap is also found in the reflected entropy of the three-boundary wormholes \cite{Hayden:2021gno}.
Referring to Fig. (\ref{fig:worm puri}), the left panel describes a three-boundary wormhole.
The black circles denote the three asymptotic
boundaries. The blue circles $L_{A}$, $L_{B}$, $L_{C}$ are geodesics of length $\sigma$.
The reflected entropy $S_{R}\left(A:B\right)$
for the bipartite mixed state $\left(A,B,C\right)$ can be calculated by canonical purification.
The process of canonical purification is similar to that of CFT: 1)
Prepare a copy of the Y-piece, say $\left(A^{*},B^{*},C\right)$;
2) Glue these two Y-pieces along the blue geodesic  $L_{C}$,
which denotes to trace out $C$; 3) The reflected surface is the minimal
geodesic between $AA^{*}$ and $BB^{*}$,  $S_{R}\left(A:B\right)=S\left(AA^{*}\right)_{\sqrt{\rho_{AB}}}$.

Looking at the right panel of Fig. (\ref{fig:worm puri}),
there are three  reflected surface candidates separating $AA^{*}$ from
$BB^{*}$. Since $\left(L_{A}+L_{A}^{\prime}\right)=\left(L_{B}+L_{B}^{\prime}\right)=2\sigma \leq2L_{*}$
belongs to the
disconnected phase, the only choice is $S_{R}\left(A:B\right)=\delta$.
Therefore, the minimal value of the reflected entropy between $A$ and
$B$ of the three-boundary wormhole is $\delta_{min}=4\mathrm{arcsinh}\left(2\sqrt{2}\right)>2L_{*}$.
One can refer to \cite{Akers:2019nfi,Li:2020ceg} for further details of
the wormhole reflected entropies.

So, the existence of the gap is not occasional. In order to understand the cause of this gap, careful studies on the
(generalized) reflected entropy from CFT perspective is necessary  in future works.

\vspace*{1.0ex}
\uline{Example $2$}
\vspace*{1.0ex}

The second example, referring to Fig. (\ref{fig:tri to ver}), is to show how to build the bulk geometry of a
tripartite system by string vertices. The classical construction is trivial: it is simply a Y-piece. Quantum construction
needs a little work and has the following steps:

\begin{figure}[H]
\begin{centering}
\includegraphics[scale=0.7]{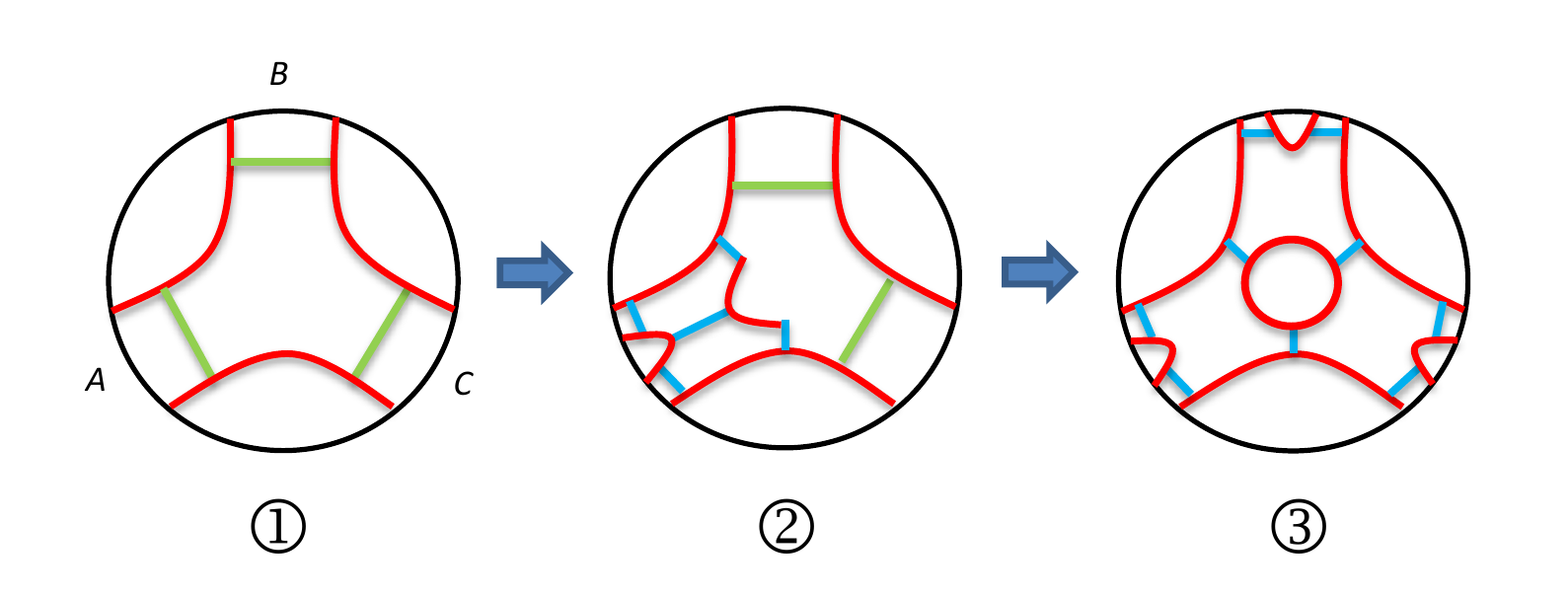}
\par\end{centering}
\caption{\label{fig:tri to ver}Rebuild tripartite canonical purification by
string vertices $\tilde{\mathcal{V}}_{1,6}$. Note all green or blue
geodesics have the same lengths. }
\end{figure}

\begin{enumerate}
\item Since the boundaries (blue) of the  string vertices have  equal length, the corresponding
bipartite reflected entropies (green)  have the same value
$S_{R}\left(A:BC\right)=S_{R}\left(B:AC\right)=S_{R}\left(B:AC\right)=s$.

\item From example 1,  the bulk geometry of
a bipartite entanglement wedge can be replaced by a string vertex $\tilde{\mathcal{V}}_{0,4}\left(L\right)$,
as illustrated by the middle panel.

\item Do the replacement for all three bipartite systems,
we  get a string vertex $\tilde{\mathcal{V}}_{1,6}\left(L\right)$
by gluing three X-pieces, or equivalently speaking, six Y-pieces
$\tilde{\mathcal{V}}_{0,3}\left(L\right)$.
\end{enumerate}
Moreover, the string vertices
$\mathcal{V}_{1,6}\left(L\right)$ and reflected entropy are mutually computable:
\begin{eqnarray}
{\color{green}S_{R}\left(\bullet:\bullet\bullet\right)} & = & {\color{green}s}=4\mathrm{arc}\sinh\frac{\cosh\frac{{\color{blue}L}}{2}}{\sinh\frac{{\color{blue}L}}{4}},\nonumber \\
\mathcal{V}_{1,6}\left({\color{blue}L}\right) & = & \mathcal{V}_{1,6}\left(4\mathrm{arc}\sinh\left[\frac{1}{4}\left(\sinh\frac{{\color{green}s}}{4}-\sqrt{\sinh^{2}\frac{{\color{green}s}}{4}-8}\right)\right]\right).
\end{eqnarray}

\vspace*{1.0ex}
\uline{Example $3$}
\vspace*{1.0ex}

\begin{figure}[H]
\begin{centering}
\includegraphics[scale=0.7]{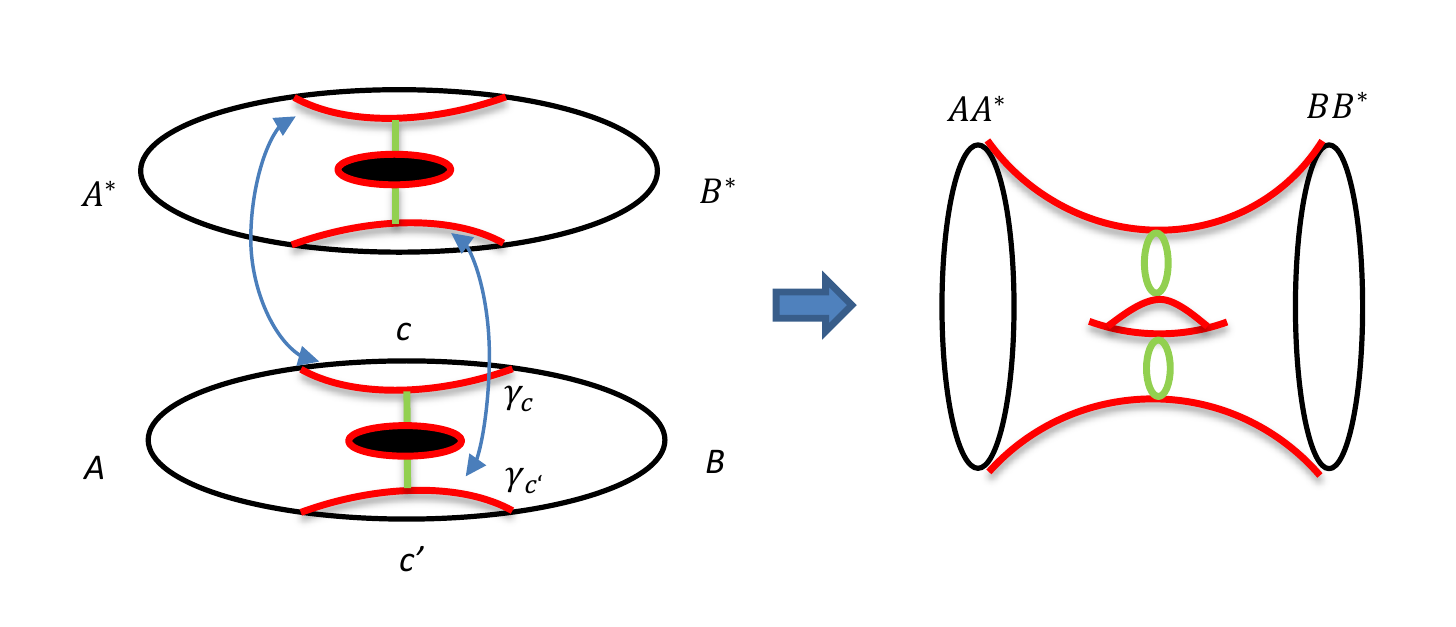}
\par\end{centering}
\caption{\label{fig:thermal puri}Canonical purification for a thermal state.}
\end{figure}

The third example is a bipartite system for a thermal state of CFT$_2$. Unlike the previous two
examples, the bulk dual of  this state intrinsically has a
black hole located at the center of the Poincare disk.
The black hole    introduces a genus after purification,
see Fig. (\ref{fig:thermal puri}), which invalidates  classical constructions.
So, this reflected surface can only be constructed quantum mechanically.

\noindent To construct the bulk geometry with string vertices, we  have the following
steps as illustrated in Fig. (\ref{fig:bi corres hole}):

\begin{figure}[H]
\begin{centering}
\includegraphics[scale=0.7]{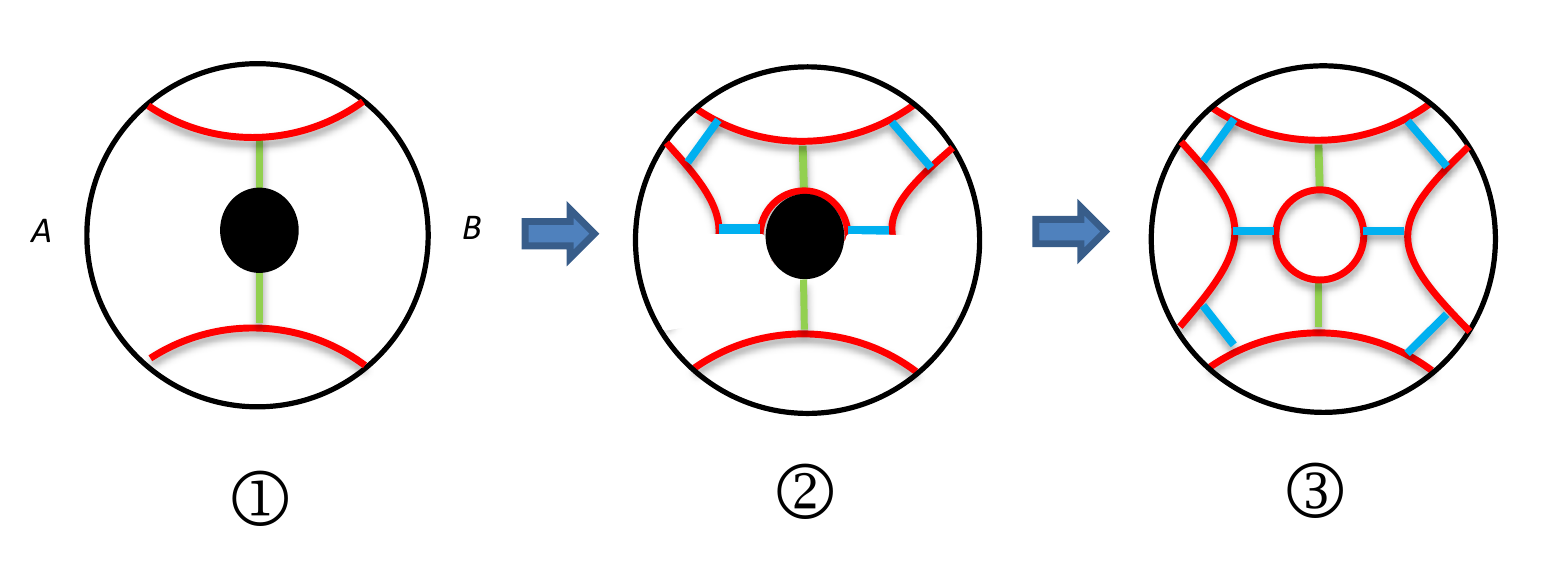}
\par\end{centering}
\caption{\label{fig:bi corres hole}Bipartite canonical purification with a
black hole can be built by $\mathcal{V}_{1,4}$.}
\end{figure}

\begin{enumerate}
\item Two disconnected green geodesics in the left panel
has the total length $S_{R}\left(A:B\right)=2s$.
\item From example $1$, we can replace each green geodesic
by a string vertices $\tilde{\mathcal{V}}_{0,4}\left(L\right)$.
\item Gluing two $\tilde{\mathcal{V}}_{0,4}\left(L\right)$ along blue boundaries,
we  see that the bulk geometry can be built by a  string
vertex $\mathcal{V}_{1,4}\left(L\right)$.
\end{enumerate}
From eq. (\ref{eq: Colollary 5}), we have

\begin{eqnarray}
{\color{green}S_{R}\left(A:B\right)} & = & 2{\color{green}s}=8\mathrm{arc}\sinh\frac{\cosh\frac{{\color{blue}L}}{2}}{\sinh\frac{{\color{blue}L}}{4}},\nonumber \\
\mathcal{V}_{1,4}\left({\color{blue}L}\right) & = & \mathcal{V}_{1,4}\left(4\mathrm{arc}\sinh\left[\frac{1}{4}\left(\sinh\frac{{\color{green}s}}{4}-\sqrt{\sinh^{2}\frac{{\color{green}s}}{4}-8}\right)\right]\right).
\end{eqnarray}

\noindent Moreover, by using eq. (\ref{eq: boundary length}),
the area $A$ of the black hole's horizon also can be fixed by the
string vertices:

\begin{equation}
A=2\mathrm{arc}\cosh\frac{\cosh\frac{L}{2}}{\cosh\frac{L}{2}-1}.
\end{equation}

\subsection{Remove the equal length restriction}

\begin{figure}[H]
\begin{centering}
\includegraphics[scale=0.7]{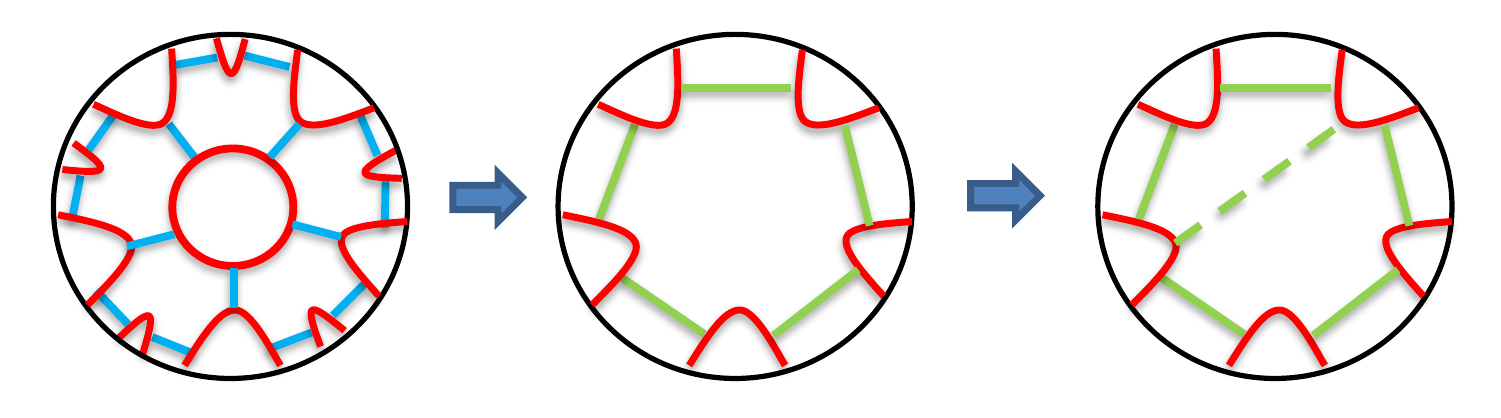}
\par\end{centering}
\caption{\small{ \label{fig:5 partite} The blue geodesics disappear when we calculate the entanglement wedge
cross sections since $L<L_{*}$. In addition to the near boundary bipartite reflected  surfaces $S_{R}\left(\bullet:\bullet\bullet\bullet\bullet\right)$ (solid green lines),
a new class of bipartite reflected
entropies (green dashed line) emerges $S_{R}\left(\bullet\bullet:\bullet\bullet\bullet\right)\neq S_{R}\left(\bullet:\bullet\bullet\bullet\bullet\right)$.}}
\end{figure}

Since the boundary geodesics of string vertex $\mathcal{\tilde  V}_{0,3}(L)$ has equal length $L<L_*$,
one may worry  the reflected surfaces built on it are too restricted.
To clarify this problem,
we  consider the canonical purification of a pentapartite
mixed state.
The dual bulk geometry can be rebuilt by  five X-pieces,
as sketched in Fig. (\ref{fig:5 partite}).
Since the geodesic lengths of string vertices have the same value $L$, all
bipartite reflected entropies $S_{R}\left(\bullet:\bullet\bullet\bullet\bullet\right)$
(solid green lines)
near the boundary are identical.
Remarkably, there emerges a new class of  bipartite reflected
entropies (green dashed line) which have different values $S_{R}\left(\bullet\bullet:\bullet\bullet\bullet\right)\neq S_{R}\left(\bullet:\bullet\bullet\bullet\bullet\right)$.

\begin{figure}[H]
\begin{centering}
\includegraphics[scale=0.3]{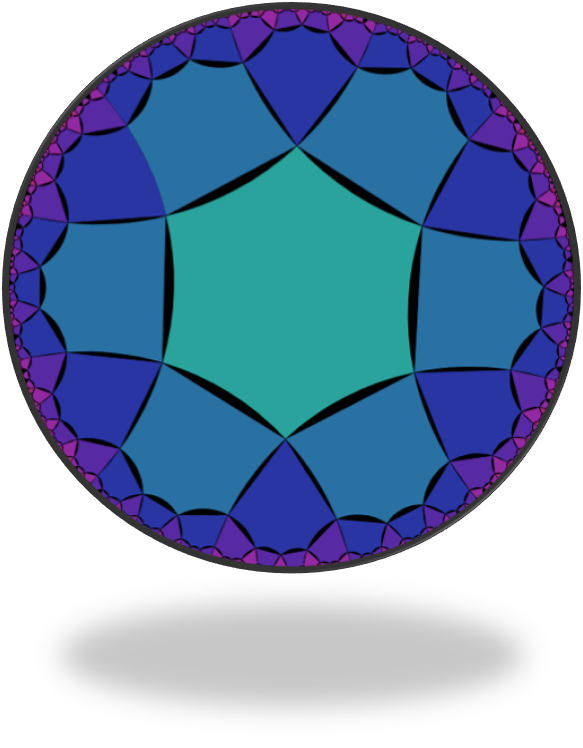}
\par\end{centering}
\caption{\small{\label{fig:tiling} The sides of all hexagons are  equal. By prolonging the
sides to the boundary, each hexagon corresponds to a different region on the boundary.
One thus can get different reflected entropies by selecting different hexagons.}}
\end{figure}

\noindent It is easy to understand that by adding more X-pieces, various reflected surfaces can be constructed.
Particularly, one   general situation is exhibited by  Fig. (\ref{fig:tiling}), where
all pieces
are right-angled equilateral hexagons. The sides of these hexagons are all equal \cite{tiling}.
Obviously, each hexagon relates
different disk boundary regions when we prolong the sides
to the boundary of the Poincare disk.
By selecting appropriate hexagons,
one can get, say, arbitrary bipartite surfaces.

\noindent Before closing this section, let us give some remarks:

\vspace*{1.0ex}

\noindent \uline{Holographic realization of phase transition}

\begin{figure}[H]
\begin{centering}
\includegraphics[scale=0.7]{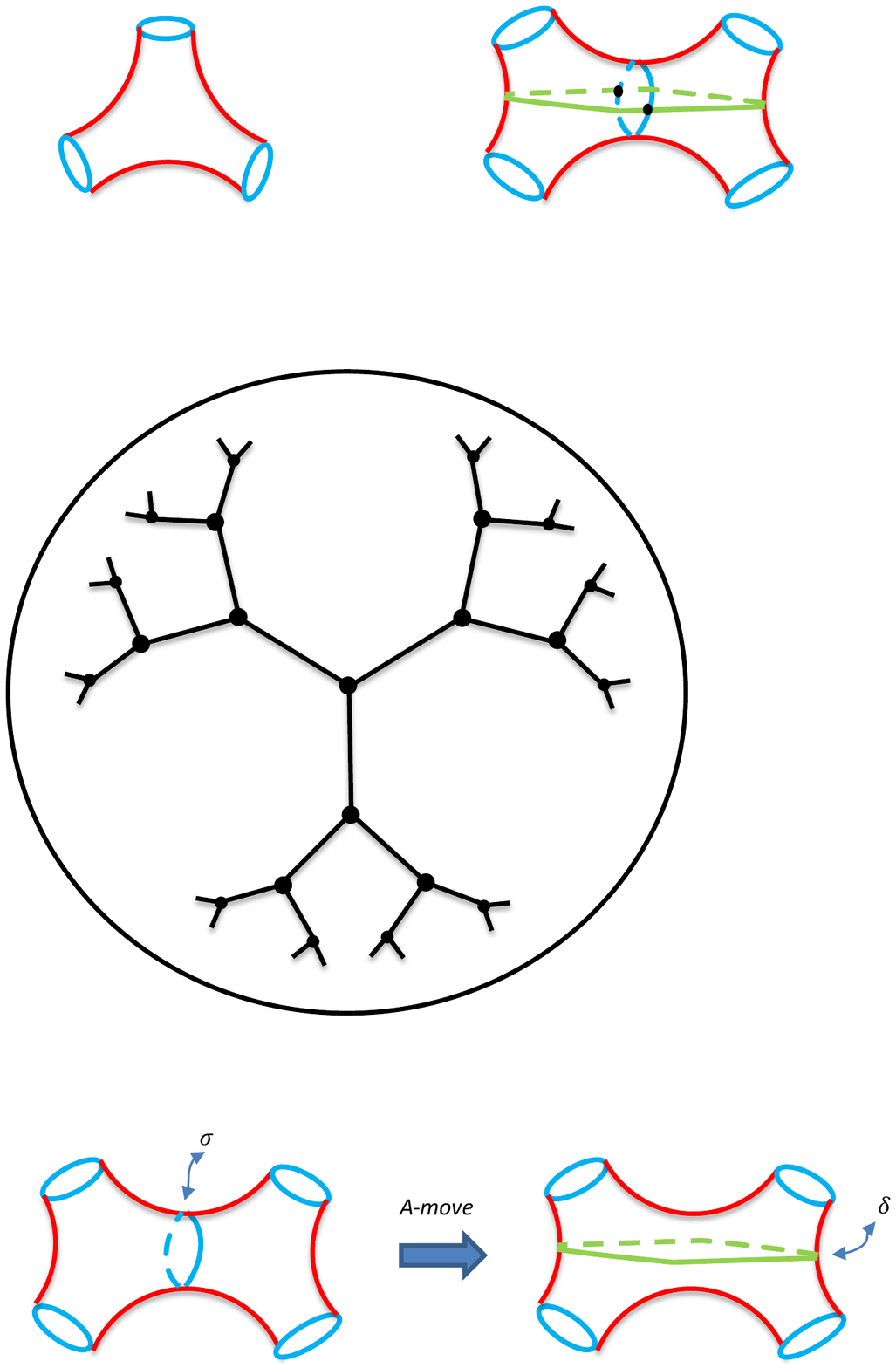}
\par\end{centering}
\caption{\label{fig:A move}A-move of the X-piece.}
\end{figure}

We have mentioned that there is a discontinuity of $S_{R}\left(A:B\right)$ at $2L_{*}$, where
$S_{R}\left(A:B\right)$ jumps from $2L_{*}$ to $0$ abruptly,
as shown by Fig. (\ref{fig:phase}).
Referring to the X-piece (\ref{fig:X-piece}) and eq. (\ref{eq: delta}), we can see that this process
basically corresponds to an exchange of the roles of $\sigma$ and $\delta$.
For a surface, there are different pants decompositions. For example, one can
decompose  an X-piece in an up-down or left-right way, as in Fig. (\ref{fig:A move}). These two
different decompositions indicate different entangling directions. Mathematically, this is called
an A-move \cite{Hatcher:1999}.
In other
words, the phase transition between the disconnected phase and the connected
phase is actually an A-move.

\section{Preliminary evidence}

In order to reinforce the connections between  two theories,
in this section, we present an  evidence of it.
As mentioned by Costello and Zwiebach \cite{Costello:2019fuh},
the hyperbolic metrics can also be used in the open string field theory.
In the classical limit, the three-open-string vertex can be constructed
with a hyperbolic hexagon $L/2$, $\vartheta$, $L/2$, $\vartheta$,
$L/2$, $\vartheta$ directly (See Fig. (\ref{fig:hexagon})), followed by grafting
three semi-infinite strips of width $L/2$ as the external legs on
the boundaries. It is then obvious that, the three-closed-string vertices are
obtained by   gluing  two three-open-string vertices. This is precisely
the pattern of  canonical purification.
Therefore, to close the logic chain between
entanglement theory and string field theory, there should be an inference that,  some (if not
all) properties of the closed string field theory (CSFT) can be constructed
by purification from the open string field theory (OSFT) with mixed
state:

\begin{eqnarray*}
\mathrm{EE}: & \begin{array}{c}
\mathrm{Eentanglement}\\
\mathrm{wedge\;cross\;section}\\
\left(\mathrm{mixed\;state}\right)
\end{array} & \underrightarrow{\mathrm{Purified}}\qquad\mathrm{Reflected\;surface}\\
 & \Downarrow\\
\mathrm{String}: & \begin{array}{c}
\mathrm{Open\;string\;field\;theory}\\
\left(\mathrm{mixed\;state}\right)
\end{array} & \underrightarrow{\mathrm{Purified}}\qquad\mathrm{Closed\;string\;field\;theory}
\end{eqnarray*}

\noindent
To build the
connections between classical OSFT and purification of entanglement,
there are  two points need to be considered: 1) Classically,
hyperbolic OSFT is a non-polynomial theory, therefore its vertices
do not satisfy the strict associativity \cite{Gaberdiel:1997ia}.
This implies that there exist $n\geq4$ open-string vertices  which
can be obtained by cutting the closed string vertices $\mathcal{V}_{0,n}$
along $\vartheta$'s. 2) In the quantum version, the loop diagrams
of open strings naturally include closed strings. It is inconsistent
to consider OSFT only. We leave this construction to the future work.

The following result has
been obtained in \cite{BottaCantcheff:2015ebn}. We simply present the
main conclusions here.

In order to study how to achieve the purification in string field
theory, we first introduce the Thermofield Double (TFD) state. In
QFT, the Hamiltonian $H$ and its eigenstate $\left|n\right\rangle $
are given by

\begin{equation}
H\left|n\right\rangle =E_{n}\left|n\right\rangle .
\end{equation}

\noindent The TFD
formalism is to double the degrees of freedom of the thermal or mixed state $\rho=e^{-\beta H}$,
and then obtain a larger
pure state. The TFD state of this doubled system is defined
as

\begin{equation}
\left|TFD\right\rangle =\frac{1}{\sqrt{Z\left(\beta\right)}}\underset{n}{\sum}e^{-\beta E_{n}/2}\left|n\right\rangle _{1}\left|n\right\rangle _{2},
\end{equation}

\noindent which is a pure state in the new system. The corresponding
reduced density matrix is given by $\rho_{total}=\left|TFD\right\rangle \left\langle TFD\right|$.
It is easy to check that when the system $2$ is traced out, the reduced
density matrix becomes $\rho_{1}=\mathrm{Tr}_{2}\rho_{total}=e^{-\beta H_{1}}$
which agrees with the original thermal state. This procedure is called
the purification of the thermal state. A comprehensive introduction
can be found in  \cite{Hartman:2015} and references therein. In the following discussions,
we will denote $\left|n\right\rangle _{1}\equiv\left|n\right\rangle $
and its copy $\left|n\right\rangle _{2}\equiv\left|\tilde{n}\right\rangle $.
Using the new notations, the Hamiltonian $\hat{H}$ of the doubled
system can be defined as:

\begin{equation}
\hat{H}\equiv H-\tilde{H}.
\end{equation}

\noindent In addition, the TFD formalism allows us to obtain the operators
$\tilde{X}$, $\tilde{Y}$,... from $X$, $Y$,... by the tilde conjugation
rules \cite{Santana}:

\begin{eqnarray}
\left(XY\right)^{\sim} & = & \tilde{X}\tilde{Y},\nonumber \\
\left(cX+Y\right)^{\sim} & = & c^{*}\tilde{X}+\tilde{Y},\nonumber \\
\left(X^{\dagger}\right)^{\sim} & = & \tilde{X}^{\dagger},\nonumber \\
\left[\tilde{X},Y\right] & = & 0,\nonumber \\
\left(\tilde{X}\right)^{\sim} & = & \epsilon X,
\end{eqnarray}

\noindent where $c^{*}$ relates to the $c^{*}$ -algebra and $\epsilon=+1\left(-1\right)$
for commuting (anti-commuting) fields.

Now, consider the Witten's cubic OSFT action \cite{Witten:1985cc},

\begin{equation}
S\left[\Phi\right]=\frac{1}{2}\left\langle \Phi,Q_{B}\Phi\right\rangle +\frac{g}{3}\left\langle \Phi,\Phi,\Phi\right\rangle ,\label{eq:OSFT action}
\end{equation}

\noindent where $Q_{B}$ is the BRST operator, $g$ is the open string
coupling constant, and $\left\langle \cdot,\cdot\right\rangle $ and
$\left\langle \cdot,\cdot,\cdot\right\rangle $ are 2-point and 3-point
vertices. The state of open string field can be expanded in the Fock
space basis:

\begin{equation}
\left|\Phi\right\rangle =\int\frac{d^{D}k}{\left(2\pi\right)^{D}}\left(T\left(k\right)+A_{\mu}\left(k\right)\alpha_{-1}^{\mu}+\cdots\right)\left|\Omega\right\rangle ,
\end{equation}

\noindent where $T$ denotes the tachyon field, $A_{\mu}$ denotes
the photon field, $\left|\Omega\right\rangle =c_{1}\left|0;k\right\rangle $
is the vacuum of the Fock space, and $\alpha_{-1}^{\mu}$ is the Fourier
mode of a string $X^{\mu}\left(\tau,\sigma\right)$. To calculate
the bracket in \eqref{eq:OSFT action}, it is useful to define the
reflector state. Using $\left|\Phi_{i}\right\rangle \in\mathcal{H}_{CFT}$
to denote a basis for states and its dual $\left\langle \Phi^{i}\right|\in\mathcal{H}_{CFT}^{*}$
such that $\left.\left\langle \Phi^{i}\right|\Phi_{j}\right\rangle =\delta_{j}^{i}$,
the correlation is given by

\begin{equation}
\left\langle \Phi_{i},\Phi_{j}\right\rangle \equiv\left.\left\langle R_{12}\right|\Phi_{i}\right\rangle _{\left(1\right)}\left|\Phi_{j}\right\rangle _{\left(2\right)},
\end{equation}

\noindent where $\left\langle R_{12}\right|\in\mathcal{H}_{CFT}^{*}\otimes\mathcal{H}_{CFT}^{*}$.

We now show  how OSFT relates to CSFT after
using the TFD formalism. In this formalism, the original one-open
string vacuum state is doubled:

\begin{equation}
\left|\Omega\right\rangle =c_{1}\left|0;k\right\rangle ,\qquad\left|\tilde{\Omega}\right\rangle =\tilde{c}_{1}\left|\tilde{0};\tilde{k}\right\rangle .
\end{equation}

\noindent Using these states, the total vacuum state of the doubled
system is defined as

\begin{equation}
\left|\left.\Omega\right\rangle \right\rangle \equiv\left|\Omega\right\rangle \varotimes\left|\tilde{\Omega}\right\rangle =c_{1}\left|0;k\right\rangle \varotimes\tilde{c}_{1}\left|\tilde{0};-k\right\rangle ,
\end{equation}

\noindent where $\tilde{k}=-k$ comes from the requirement that the
purified state $\left|\left.\Omega\right\rangle \right\rangle $ must
equal  its tilde conjugate: $\widetilde{\left|\left.\Omega\right\rangle \right\rangle }=\left|\left.\Omega\right\rangle \right\rangle $.
We also have

\begin{equation}
\left(L_{0}-1\right)\left|\left.\Omega\right\rangle \right\rangle =0,\qquad\left(\tilde{L}_{0}-1\right)\left|\left.\Omega\right\rangle \right\rangle ,\qquad\left(L_{0}-\tilde{L}_{0}\right)\left|\left.\Omega\right\rangle \right\rangle =0.
\end{equation}

\noindent Based on the definition of the vacuum state, the states of
open string field are doubled:

\begin{eqnarray}
\left|\left.\Phi\right\rangle \right\rangle  & = & \int\frac{d^{D}k}{\left(2\pi\right)^{D}}\left(T\left(k\right)+A_{\mu}\left(k\right)\alpha_{-1}^{\mu}+B_{\mu}
\left(k\right)\tilde{\alpha}_{-1}^{\mu}\cdots\right)\left|\left.\Omega\right\rangle \right\rangle ,\nonumber \\
\left|\left.\tilde{\Phi}\right\rangle \right\rangle  & = & \int\frac{d^{D}k}{\left(2\pi\right)^{D}}\left(\tilde{T}\left(k\right)+\tilde{A}_{\mu}\left(k\right)\tilde{\alpha}_{-1}^{\mu}+\tilde{B}_{\mu}
\left(k\right)\alpha_{-1}^{\mu}\cdots\right)\left|\left.\Omega\right\rangle \right\rangle .
\end{eqnarray}

\noindent Therefore, the kinetic term of OSFT can be extended with the TFD
formalism to:

\begin{equation}
\hat{S}\left[\Phi,\tilde{\Phi}\right]=S\left[\Phi\right]-\tilde{S}\left[\tilde{\Phi}\right].\label{eq:TFD action}
\end{equation}

\noindent The corresponding reflector   is  $\left\langle \left\langle R_{12}\right.\right|\equiv\left\langle R_{12}\right|\varotimes\left\langle R_{12}\right|$
which also equals its tilde conjugate. More explicitly, the free-OSFT action
is

\begin{equation}
\hat{S}\left[\Phi,\tilde{\Phi}\right]=\frac{1}{2}\left\langle \left\langle \Phi,Q_{B}\Phi\right\rangle \right\rangle -\frac{1}{2}\left\langle \left\langle \tilde{\Phi},\tilde{Q}_{B}\tilde{\Phi}\right\rangle \right\rangle ,\label{eq:free-OSFT action}
\end{equation}

\noindent with   equations of motion:

\begin{equation}
Q_{B}\left|\left.\Phi\right\rangle \right\rangle =0,\qquad\tilde{Q}_{B}\left|\left.\tilde{\Phi}\right\rangle \right\rangle =0,\label{eq:EOM}
\end{equation}

\noindent which are the same as the physical condition in the BRST
quantization. To relates this action \eqref{eq:free-OSFT action}
with the action of CSFT, one introduces the generally
entangled ground states, namely $\Omega$-states, which is defined by

\begin{equation}
\left(L_{0}-\tilde{L}_{0}\right)\left|\left.\Omega\left(\theta\right)\right\rangle \right\rangle =0,\qquad\left|\left.\Omega\left(\theta\right)\right\rangle \right\rangle =\widetilde{\left|\left.\Omega\left(\theta\right)\right\rangle \right\rangle },\label{eq:level-matching}
\end{equation}

\noindent where $\theta$ denotes the level of a component field which
characterizes the vacuum. Using $\Omega$-states, the equations of
motion of \eqref{eq:free-OSFT action} give

\begin{equation}
Q_{B}\pm\tilde{Q}_{B}\left|\left.\Omega\left(\theta\right)\right\rangle \right\rangle =0.\label{eq:new EOM}
\end{equation}

\noindent The  ``$+$'' sign gives the equations of motion of
free-CSFT and the first equation of \eqref{eq:level-matching} is precisely
the level-matching condition. Moreover,
$\left|\left.\Omega\left(\theta\right)\right\rangle \right\rangle $
also can be expanded in $\left|\left.\Omega\right\rangle \right\rangle $:

\begin{equation}
\left|\left.\Omega\left(\theta\right)\right\rangle \right\rangle =\int\frac{d^{D}k}{\left(2\pi\right)^{D}}\left(T\left(k\right)+C_{\mu\nu}\left(k\right)\alpha_{-1}^{\mu}\tilde{\alpha}_{-1}^{\nu}\cdots\right)\left|\left.\Omega\right\rangle \right\rangle ,
\end{equation}

\noindent where $t\left(k\right)=\tilde{t}\left(k\right)$, $C_{\mu\nu}\left(k\right)=\tilde{C}_{\nu\mu}\left(k\right)$,
and $\theta$ denotes the level of the component field $\left(t,C_{\mu\nu},\ldots\right)$.
$C_{\mu\nu}$ can be decomposed into $g_{\mu\nu}+b_{\mu\nu}+\phi\eta_{\mu\nu}$ to
include the closed string massless sector. In addition, from the equations
of motion \eqref{eq:new EOM}, it is easy to figure out the action:

\begin{equation}
\hat{S}_{\mathrm{Canonical}}\left[\Omega\right]=\frac{1}{2}\left(\left.\left\langle \left\langle R_{12}\right|\Omega\right\rangle \right\rangle \right)\hat{c}_{0}\left(Q_{B}+\tilde{Q}_{B}\right)\left|\left.\Omega\right\rangle \right\rangle,
\end{equation}

\noindent which is the free-CSFT action. Therefore, $\Omega$-states
can be seen as the asymptotic states of free-CSFT. For the higher order interacting
vertices, the derivations are not easy.
As an alternative, one
can check the S-matrix and expect the results to agree with the closed
string. The detailed discussions can be found in \cite{BottaCantcheff:2015ebn}.

\vspace*{4.0ex}

\section{Hyperbolic closed string vertices as spacetime building blocks?}

In the previous discussions, we have proposed a connection between
the reflected entropy/surfaces and hyperbolic string vertices. Since
it is widely believed that the spacetime structure could be generated
by the entanglement entropy through the dual surfaces, it is natural
to ask if the spacetime structure could directly emerge from the hyperbolic
string vertices. The advantage of the hyperbolic string vertices approach
is that we have a dynamical equation, the BV master equation, to control
the generating process.

\begin{figure}[H]
\begin{centering}
\includegraphics[scale=0.7]{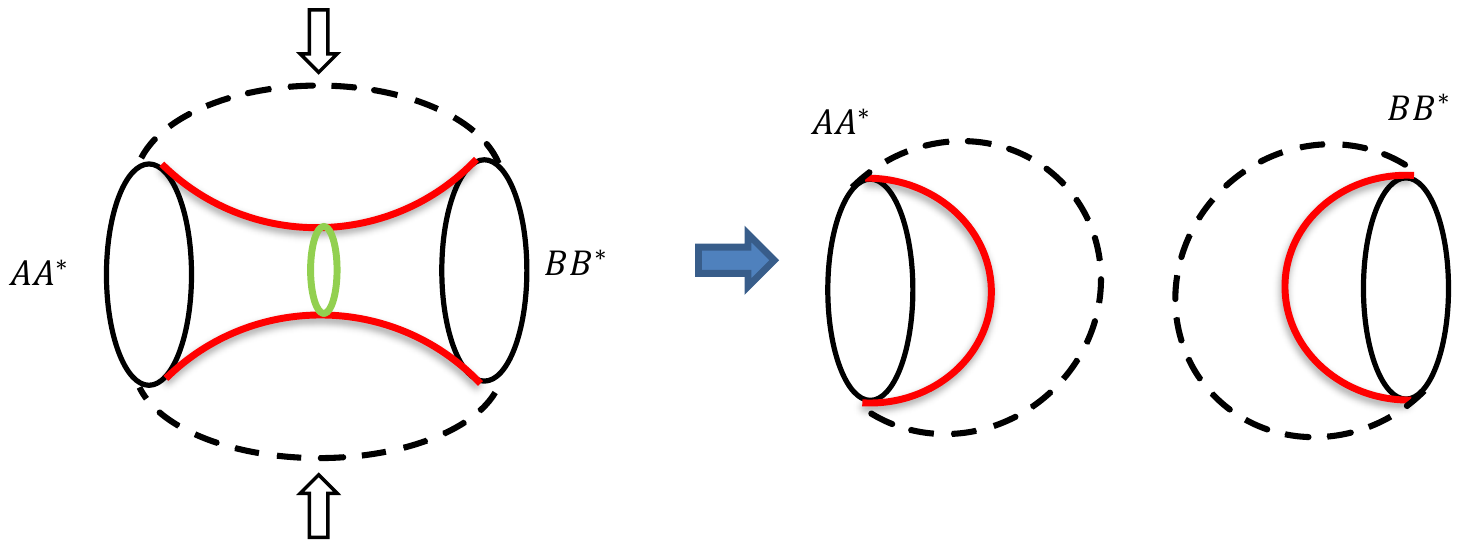}
\par\end{centering}
\caption{\label{fig:building 1} Disconnect a bipartite mixed state into two
separate parts in the picture of entanglement entropy.}
\end{figure}

To explain this idea more explicitly, let us consider the disentangling
process of a bipartite system in Fig. (\ref{fig:building 1}). In
this mixed state, once regions $A$ and $B$ are fixed, the reflected
surface $S_{R}\left(A:B\right)>2L_{*}$ is also fixed. How to disentangle
regions $A$ and $B$ without changing the boundary geometry? This
is different from the phase transition discussed in section $2.2$
where regions $A$ and $B$ are changed. Following the arguments about
emergent spacetime introduced in \cite{VanRaamsdonk:2010pw} , the
only way to realize the disentanglement is to disconnect the spacetime
into two separate parts.

\begin{figure}[H]
\begin{centering}
\includegraphics[scale=0.7]{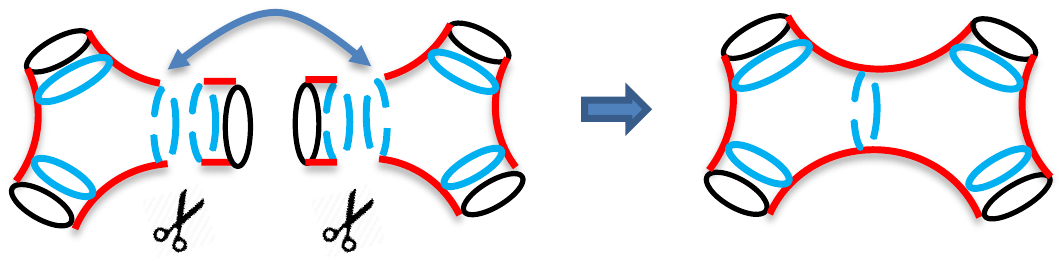}
\par\end{centering}
\caption{\label{fig:BV1-1}Creation of a baby universe by gluing two hyperbolic
string vertices $\mathcal{V}_{0,3}$.}
\end{figure}

On the other hand, we can consider this process from the perspective
of hyperbolic closed string vertices, let us focus on the simplest
case: 2D universes. It is verified and well understood that the 2D
universes can be described by string theory. The sum-over-two-geometries
of string field theory can be considered as a theory of splitting
and joining of 2D baby universes \cite{Cohen:1985sm,Giddings:1988wv}.
Therefore, the hyperbolic string vertices of CSFT indeed play the
role of spacetime building blocks of interacting 2D universes. In
Fig. (\ref{fig:BV1-1}) we present how a joining of universes can
be viewed as the BV gluing process of hyperbolic string vertices $\mathcal{V}_{0,3}$:
we remove two disks of each three-sphere by cutting along the
blue geodesics and then glue them together. This operation is also
discussed in holographic baby universes in ref. \cite{Gesteau:2020wrk}.
The process is controlled by the BV equation, and the only way to
destroy the entanglement is to disconnect the spacetime into two separate
parts, which is triggered by tachyon condensation \cite{Adams:2005rb}.

\begin{figure}[H]
\begin{centering}
\includegraphics[scale=0.8]{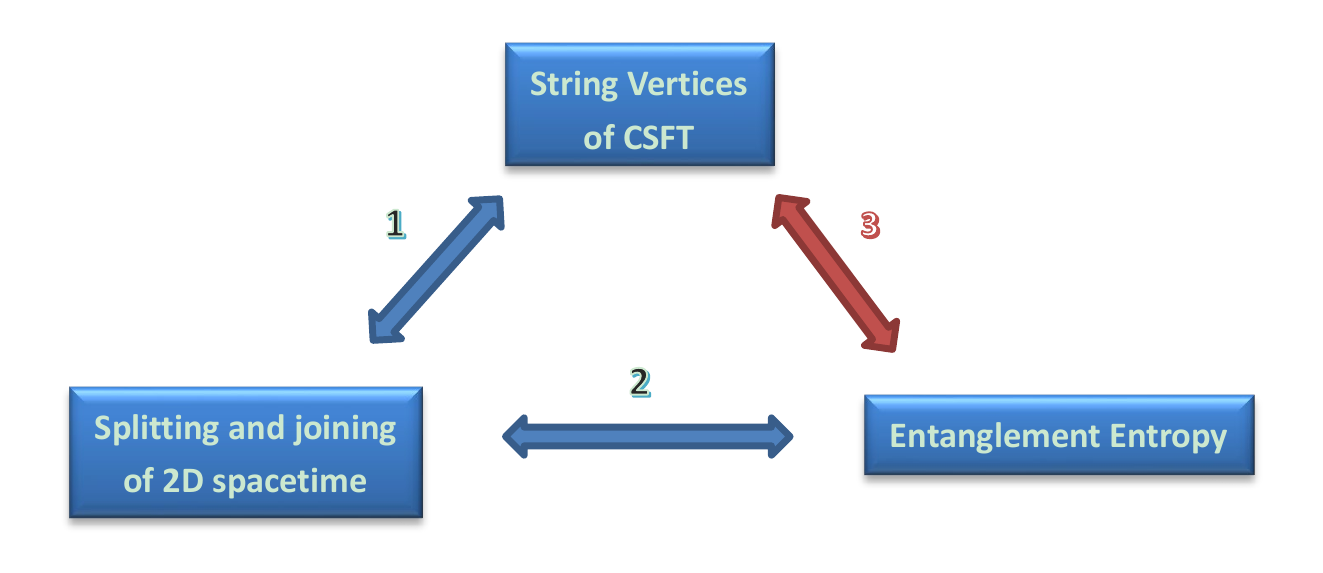}
\par\end{centering}
\caption{\label{fig:triangle}Triangular relation: 1. Giddings and Strominger's
baby universes. 2. Van Raamsdonk's conjecture. 3. Connections between
reflected entropies and hyperbolic string vertices.}
\end{figure}

Based on this observation, we can complete a triangular relation as depicted in Fig. (\ref{fig:triangle}).
First, it is known that the string vertices correspond to the splitting
and joining of 2D baby universes, which is proposed by Giddings
and Strominger ($1$ in Fig. (\ref{fig:triangle})) \cite{Giddings:1988wv}. Then, Raamsdonk
argues that the splitting and joining of spacetime relate to the quantum
entanglement ($2$ in Fig. (\ref{fig:triangle})) \cite{VanRaamsdonk:2010pw}. Finally, our discussions
of the connections between reflected entropies and hyperbolic string
vertices complete this triangle ($3$ in Fig. (\ref{fig:triangle})).

Then, it is reasonable to conjecture that   higher  dimensional $D>2$ spacetime
also can be affected by the interactions of closed strings.
Of course, there are many  works need be done to verify this
point. As a result,\emph{ spacetime tells worldsheet how to move,
and worldsheet tells spacetime how to glue}.

\section{Conclusion and discussions}

In short, we proposed close connections between the reflected entropies/surfaces
of the multipartite mixed states of CFT$_{2}$ and the hyperbolic
string vertices of CSFT. We found that the reflected entropies could be determined
by hyperbolic string vertices and vice versa. We consequently conjectured that the hyperbolic
string vertices can play a major role in   building
spacetime.

Some remarks and future works are as follows.
\begin{itemize}
\item Cho has generalized constructions of the closed string vertices
to the open-closed string vertices \cite{Cho:2019anu}.
It is worth extending our results to find the boundary dual of the open-closed string vertices.
Are they some new entanglement quantities or existing ones?

\item In subsection 3.1, we explained what accounts for the gap
between $\delta_{min}$ and   $S_{Rmin}$.
Note that $\delta_{min}$ is for the \emph{geodesic}
bounded hyperbolic surfaces, which are precisely
the reflected surfaces or the hyperbolic string vertices.
Nevertheless, $S_{Rmin}$ is for the  \emph{horocycle} bounded hyperbolic
surfaces which are the Moosavian-Pius
surfaces. Therefore, from this point of view,
the bulk dual of the reflected entropy of CFT is not the reflected surface,
but the  Moosavian-Pius surface!

Given that Moosavian-Pius surface is a limit of the reflected surface or hyperbolic string vertex,
we are  led to believe that
there should exist a \emph{generalized} reflected entropy in the CFT which is the real dual of the reflected surface.
This \emph{generalized} reflected entropy has the reflected entropy as its limit.

\item The random stabilizer tensor networks (RSTN) are
studied in ref. \cite{Hayden:2016cfa}. Canonical purification
for the tensor networks are considered in ref. \cite{Akers:2019gcv}.
As we showed in section $5$, the tensor networks share great similarity with
the interactions of hyperbolic string vertices,
it is of interest to study their relation more carefully and deeply.

\item Geometric master equation controls various interactions (such as taking
boundary, or sewing together) between vertices $\mathcal{V}_{g,n}$. Since
$\mathcal{V}_{g,n}$ is connected to the canonical purification of
mixed states. There could be a boundary version of geometric
master equation which tells us how two systems to entangle with each
other.

\item Based on our results, it is possible to define the moduli space of entanglement entropy.
Since hyperbolic string vertices satisfy the BV equation on the moduli space,
it is possible to get  modularized equations of motion of the  CFT.

\item The connections between the canonical purification of the multipartite
mixed states of CFT$_{2}$ and the hyperbolic string vertices of CSFT
are special cases of AdS$_{3}$/CFT$_{2}$, since the closed string
vertices possess the hyperbolic geometry intrinsically.
In addition, it  indicates that AdS/CFT has been encoded in string theory, since the string amplitudes can be calculated by CFT on punctured Riemann surfaces or hyperbolic string vertices in their bulk.

\item In this paper, we are only concerned with $d=2$ case.
one might wonder if it is possible to build higher dimensional bulk geometries
with string vertices. We anticipate that higher-dimensional foundamental objects: branes,
might achieve this purpose.


\end{itemize}

\vspace{5mm}

{\bf Acknowledgements}
We are deeply indebted to Amr Ahmadain, Aron Wall and Zihan Yan for illuminating discussions. This work is supported in part by the NSFC (Grant No. 12105191, 11947225 and 11875196). HW is supported by the International Visiting Program for Excellent Young Scholars of SCU.

\end{document}